\documentclass[submission, Phys]{SciPost}
\usepackage{amsmath,braket}
\usepackage{amsthm}
\usepackage{amssymb}
\usepackage{graphicx}
\usepackage{hyperref}
\usepackage{color}
\usepackage[textsize=tiny]{todonotes}
\usepackage{tikz}
\usetikzlibrary{calc}

\usepackage{empheq}

\usetikzlibrary{positioning,shapes.geometric,decorations.markings,arrows,knots,hobby} 
\usepackage[many]{tcolorbox}
\usepackage{pgfplots}
\usepgfplotslibrary{patchplots}
\usepgfplotslibrary{fillbetween}

\usepackage{dsfont}

\theoremstyle{definition}

\theoremstyle{remark}

\definecolor{accent1}{HTML}{A65B6F}
\definecolor{light}{HTML}{EBEDF2}
\definecolor{medium}{HTML}{D2D4D9}
\definecolor{dark}{HTML}{7297A6}
\definecolor{accent2}{HTML}{A6554E}

\usepackage{amsthm}
\usepackage{float}
\usepackage{graphicx}
\usepackage{bm}
\usepackage[utf8]{inputenc}
\usepackage[english]{babel}
\usepackage{amsfonts}
\usepackage{hyperref}
\usepackage{physics}
\usepackage{epstopdf}
\usepackage{mathtools}
\usepackage[capitalise]{cleveref}
\usepackage[space]{grffile}
\usepackage{color}
\usepackage[normalem]{ulem}
\usepackage{subcaption}
\usepackage{subcaption}
\begin{document}

\begin{center}{\Large \textbf{
Generalized hydrodynamics of free fermions under extensive-charge monitoring
}}\end{center}

\begin{center}
Pablo Bayona-Pena\textsuperscript{1*},
Michele Mazzoni\textsuperscript{1},
Lorenzo Piroli\textsuperscript{1}
\end{center}

\begin{center}
{\bf 1} Dipartimento di Fisica e Astronomia, Universit\`a di Bologna and INFN,
Sezione di Bologna, via Irnerio 46, 40126 Bologna, Italy
\\

* juanpablo.bayonapen2@unibo.it
\end{center}

\begin{center}
\today
\end{center}

% For convenience during refereeing: line numbers
%\linenumbers

\section*{Abstract}
{\bf
We study transport dynamics of free fermions subject to the external monitoring of a conserved charge over an extensive region. Focusing on bipartition protocols, we consider monitoring the total particle number over half of the system, and study the profiles of local charges and currents at hydrodynamic scales. While the Lindbladian describing the averaged dynamics is non-local, we show that the profiles can be understood in terms of localized impurities. We present a general framework based on the generalized hydrodynamics (GHD) picture, allowing for a hybrid numerical-analytic solution of the quench dynamics at hydrodynamic scales. We illustrate our approach for domain-wall initial states, showing that monitoring leads to discontinuities in the profiles that become more pronounced as the rate increases and that lead to the absence of transport in the Zeno limit of infinite monitoring rates. Our GHD framework could be naturally extended to interacting systems, paving the way for a systematic study of transport of integrable models subject to extensive-charge measurements.}
%We also study numerically the case in which the system is initialized in a homogeneous thermal state, showing that monitoring leads to a variation of the hydrodynamic profiles at the accessible time scales. }

\vspace{10pt}
\noindent\rule{\textwidth}{1pt}
\tableofcontents\thispagestyle{fancy}
\noindent\rule{\textwidth}{1pt}
\vspace{10pt}

\section{Introduction}

Monitoring a many-body quantum system may have drastic effects on its dynamics, leading to a variety of collective phenomena that go beyond those observed in isolated systems. For instance, it has been known for a long time that the presence of monitoring or dissipative effects can qualitatively change the transport properties of a quantum system~\cite{prosen2009matrixproductsimulations,znidaric2010exact,znidaric2010dephasinginduceddiffusivetransport, prosen2012diffusivehightemperaturetransport, znidaric2013transportina ,bertini2015macroscopic,bertini2021finitetemperature}. For another example, a recent important discovery has been the existence of exotic types of measurement-induced phase transitions characterizing the scaling of bipartite entanglement in unitary dynamics interspersed by local measurements for brickwork quantum circuits~\cite{li2018quantumzenoeffect,li2019measurementdrivenentanglement,skinner2019measurementinducedphase,fisher2023random,potter_entanglement_2022} and Hamiltonian dynamics~\cite{FujiAshida2020Measurement,AlbertonBuchholdDiehl2021MonitoredFermions,TurkeshiEtAl2021IsingMonitoring,FavaEtAl2023SigmaMonitoredFermions,XingEtAl2024InteractionsMonitoring}. In fact, the past decade has witnessed an increasing interest in the interplay between unitary evolution and the effects of monitoring, in large part motivated by the emergence of digital quantum platforms allowing for a variety of measurement protocols and monitored dynamics~\cite{Zhang_2017 , Arute_2019, zhang2022digital, noel2022measurement,koh2023measurement, evered2025probing}.

Generally speaking, external measurements introduce randomness in the dynamics, thus destroying any conservation law that the system may have~\cite{fisher2023random}. Therefore, one typically expects that the integrability of the Hamiltonian, defined by the existence of infinitely-many conservation laws~\cite{caux2011remarks}, does not manifest itself in any of the features of the corresponding monitored dynamics. While this is often the case, one can still consider fine-tuned measurement protocols that preserve some features of integrability in the (averaged) monitored dynamics, possibly leading to an integrable Lindbladian description~\cite{medvedyeva2016Exact,rowlands2018noisy,naoyuki2019dissipative,naoyuki2019dissipativespin,ziolkowska2020yang,essler2020integrability,buvca2020bethe,nakagawa2021exact}. A motivation to study these protocols lies in the peculiar features of integrable systems out of equilibrium~\cite{calabrese2016introduction}, raising the question of whether the interplay between integrability and external monitoring may give rise to novel many-body effects. 

In this work, we focus on transport properties of integrable systems and study special types of measurement protocols that preserve integrability at large space-time scales. Transport in integrable models is by now a mature topic~\cite{alba2021generalized,denardis2022correlation}. In this context, an important milestone has been the introduction of generalized hydrodynamics (GHD)~\cite{CastroAlvaredoDoyonYoshimura2016GHD,Bertin2017ghd}, a powerful theory that allows one to describe integrable dynamics in generally inhomogeneous settings. GHD was initially developed in the study of so-called bipartition protocols~\cite{alba2021generalized,Bertin2017ghd,piroli2017transport},  which are quantum quenches where the initial state is obtained by joining together two different homogeneous states. While conceptually very simple, this setting allows one to study a variety of exotic phenomena, from super-diffusion~\cite{ljubotina2017spin,superdiffusion2018}, to universal dynamical features~\cite{bernard2015non,bertini2018universal}, and spin-charge separation effects~\cite{mestyan2019spin,scopa2021real}. Here, we enrich the standard bipartition protocol and consider monitoring a conserved charge over a possibly extensive interval. We study the averaged dynamics and focus on the profiles of the Hamiltonian local conserved charges and the corresponding currents at large space-time scales.

We mention that the hydrodynamics of integrable systems has been already explored in the context of open quantum systems. For example, it is known that continuous monitoring of the local occupation number for particle-preserving systems yields diffusive behavior of the corresponding density profiles in the long time regime~\cite{medvedyeva2016Exact,IshiyamaFujimotoSasamoto2025ExactDensityProfile,znidaric2010exact,znidaric2014exact}, which is closely related to the physics of classical exclusion processes~\cite{IshiyamaFujimotoSasamoto2025ExactDensityProfile}. On the other hand, ballistic hydrodynamics has  been shown to appear as an effect of boundary driving~\cite{TurkeshiSchiro2021BoundaryDephasing}, localized losses \cite{AlbaCarollo2022LocalizedLosses}, or as an early time behavior that ultimately transitions to diffusive propagation~\cite{eisler2011crossover}. More recently, it has been shown that GHD allows one to quantitatively describe non-integrable Lindbladian dynamics in the limit of weak dissipation~\cite{lange2018timedependent,rossini2021strong,RossoBiellaMazza2022StrongTwoBodyLossesTrap,RiggioRossoKarevskiDubail2024HardcoreBosonsLosses,LumiaAupetitDialloDubailCollura2025AccuracyTDGGE,MarcheYoshidaNardinKatsuraMazza2026NonReciprocityTDGGE}. Our work differs from previous studies in that we explore the effects of extensive charge measurements on quantum transport in bipartite settings. 

Concretely, we focus on the simplest case of bipartition protocols in non-interacting free fermionic systems, and consider monitoring the particle number over half of the system. To the best of our knowledge, the extension of GHD to mixed-state dynamics generated by continuous monitoring of extensive charges has not been addressed before. A monitoring protocol similar to ours, however, was employed in Ref.~\cite{travaglino2025quench}, where the authors investigate bipartite entanglement entropy in a quantum quench with projective measurements of an extensive charge.

We introduce a general framework based on GHD that we exploit to obtain a hybrid numerical-analytic solution of the quench dynamics. We illustrate our approach for domain-wall initial states, showing that the monitoring causes discontinuities in the profiles that become more pronounced as the rate increases and that lead to the absence of transport in the Zeno limit of infinite monitoring rates. We also validate our predictions in the case where the system is initialized in a homogeneous thermal state, showing that monitoring leads to a variation of the hydrodynamic profiles. Our results pave the way for a systematic study of transport properties of integrable systems subject to extensive-charge measurements.

The rest of this work is organized as follows. We begin in Sec.~\ref{sec:Settings}, where we discuss the model and our measurement protocol, outlining our main results. In Sec.~\ref{sec:GHD}, we review the standard GHD approach to unitary bipartitioning protocols and extend it to the case of extensive-charge measurements. We apply this framework in Sec.~\ref{sec:micro_general}, where we present a hybrid numerical-analytic solution for the quench protocol starting from a special domain-wall initial state. Next, in Sec.~\ref{sec:thermal_states} we study the case of an initial homogeneous thermal states. Finally, our conclusions are presented in Sec.~\ref{sec:conclusions}, while the most technical parts of our work are consigned to a few appendices.

\section{The setup}
\label{sec:Settings}

\subsection{The model}

We consider a one-dimensional chain of $2N$ fermionic modes, described by the tight-binding Hamiltonian 
\begin{equation}
\hat{H} = -J\sum_{j=-N}^{N-1} \left[ \hat{c}^\dagger_j \hat{c}_{j+1} + \hat{c}^\dagger_{j+1} \hat{c}_{j}\right], \label{eq:Hamiltonian}
\end{equation}
where we assume periodic boundary conditions.
We recall that $\hat{H}$ is diagonalized by introducing the Fourier transform of the fermionic modes $\hat{c}_j$. We denote them by $\hat{\tilde{c}}_{k}$, where we introduced the quantized momenta $k=\frac{\pi}{N}j$, with integer $j\in [-N,N-1]$. Denoting by $\ket{0}$ the vacuum state, the eigenstates of $\hat H$ are Fock states of the form $\ket{k_1,\ldots ,k_n}=\hat{\tilde{c}}^\dagger_{k_1}\hat{\tilde{c}}^\dagger_{k_2}\ldots \hat{\tilde{c}}^\dagger_{k_n}|0\rangle$ and correspond to eigenvalues 
\begin{equation}
    E=\sum_{j=1}^n\varepsilon(k_j)\,,
\end{equation}
where the single-particle energy is given by
\begin{equation}
    \varepsilon(k)= -2J\cos(k)\,.\label{eq:H_esp}
\end{equation} 

The Hamiltonian~\eqref{eq:Hamiltonian} preserves the total particle number
\begin{equation}\label{eq:total_particle_num}
\hat{Q}= \sum_i \hat{c}^\dagger_i \hat{c}_i= \sum_i \hat{n}_i\,.
\end{equation}
In fact, $\hat H$ is integrable, displaying an infinite set of local conserved operators (or charges). A complete set is given by~\cite{EsslerFagotti2016Quench}
\begin{equation}
\hat{Q}^{r,\pm}=\sum_x \hat{q}^{(r,\pm)}_x,
\end{equation}
where $r\geq 0$ is an integer parameterizing the charges, while
\begin{align}
        \hat{q}^{(r,+)}_x&= J(\hat{c}^\dagger_{x}\hat{c}_{x+r} +\hat{c}^\dagger_{x+r}\hat{c}_{x})\,, \label{eq:qp}\\[3pt]
    \hat{q}^{(r,-)}_x&=i J( \hat{c}^\dagger_{x}\hat{c}_{x+r} -\hat{c}^\dagger_{x+r}\hat{c}_{x})\,. \label{eq:qm}
\end{align}
The set $\{\hat{Q}^{r,\pm}\}_{r}$ is complete in the sense that it spans the same operator space as the set of the occupation-number operators $\hat{n}(k)=\hat{\tilde{c}}^\dagger_k\hat{\tilde c}_k$~\cite{EsslerFagotti2016Quench}. 

Finally, we recall that the charges act diagonally on the eigenstates of the Hamiltonian $\ket{k_1,\ldots ,k_n}$. Namely,
\begin{equation}
 \hat{Q}^{r,\alpha}\ket{k_1,\ldots ,k_n}=\left(\sum_{j=1}^n q^{(r,\alpha)}(k_j)\right)\ket{k_1,\ldots ,k_n},   
\end{equation} 
where
\begin{align}
q^{(r,+)}(k)&= 2J\cos(rk),\, \\[3pt] 
q^{(r,-)}(k)&= -2J\sin(rk).
\end{align}

\subsection{The nonequilibrium protocol}\label{sec:the_protocol}

\begin{figure}
    \centering
\includegraphics[height=0.33\textheight,width=0.93\textwidth]{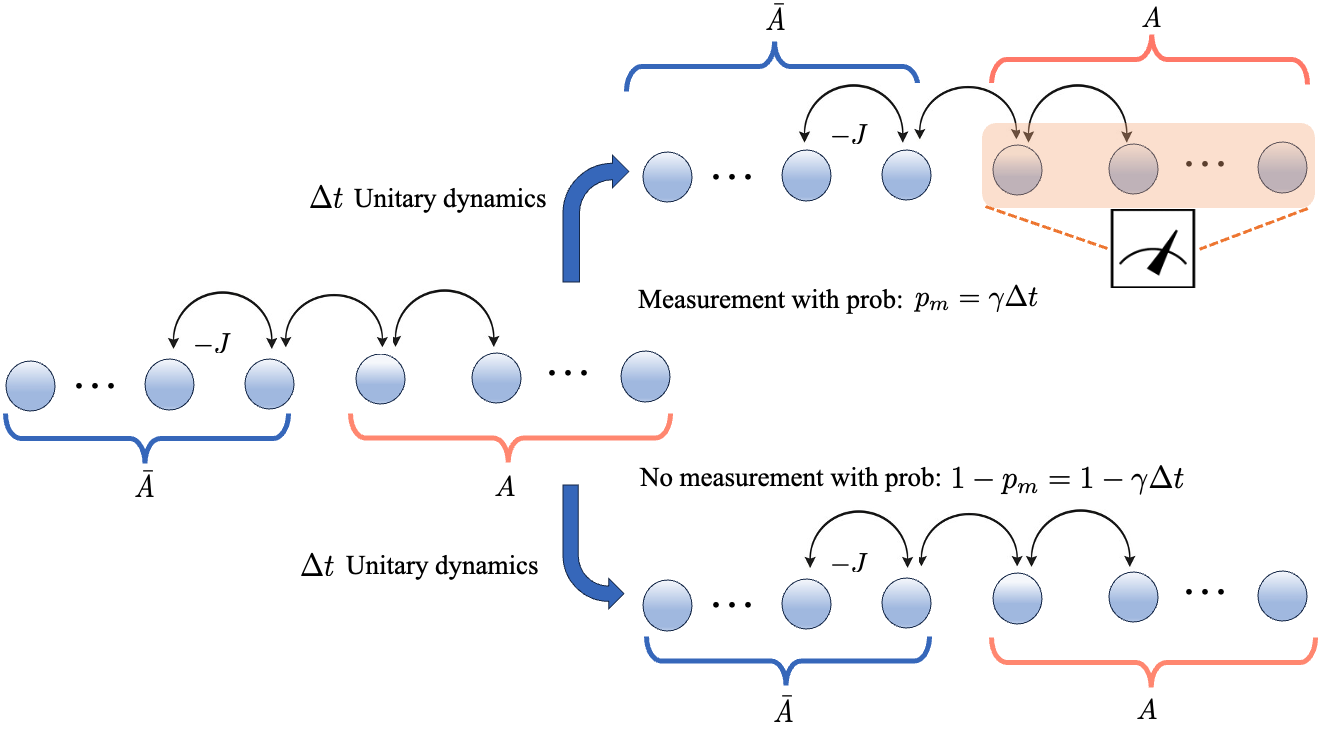}
    \caption{Pictorial representation of the monitoring protocol studied in this work. The system is initially prepared in the state $\rho_0$. Next, the system is left to evolve unitarily with Hamiltonian $\hat{H}$  and hopping rate $-J$ for a time interval $\Delta t$, followed by a incoherent measurement of the charge $\hat{Q}_A$ for the right half of the system $A=\{x \; |0<x< N\}$, with probability $p_m= \gamma \Delta t$. The dynamics of the system after averaging over trajectories is given by the Lindblad equation Eq.~\eqref{eq:Lindblad}.}
    \label{fig:Fig1_model}
\end{figure}

In this work, we will consider bipartition protocols consisting in joining together two different initial states, and will be interested in the subsequent (monitored) nonequilibrium dynamics. The protocol is explained in the following and depicted in  Fig.~\ref{fig:Fig1_model}.

First, we consider initializing the system in the state
\begin{equation}\label{eq:initial_state}
    \hat{\rho}_0=\hat\rho_L\otimes \hat\rho_R\,,
\end{equation}
where the left (right) density matrix $\hat\rho_L$ ($\hat\rho_R$) is a function of the left (right) fermionic modes $\hat c_j$, with $j\leq 0$ ($j>0)$. We will be especially interested in the case of pure states, $\hat \rho_{\alpha}=\ket{\psi_{\alpha}}\bra{\psi_{\alpha}}$, with $\alpha=L,R$.

After initialization, we let the system evolve unitarily, while continuously measuring the total particle number over an interval $A$, namely
\begin{equation}
    \hat{Q}_A=\sum_{j\in A} \hat{n}_j\,.
\end{equation}
The continuous charge monitoring can be modeled as the continuous limit of either a repeated weak-measurement process or a repeated projective-measurement process. Here, we consider the latter approach and study the following physical setting: After any small time-interval $\Delta t$, we perform a projective measurement of the charge $\hat{Q}_A$ with probability $p=\gamma \Delta t$, while we perform no measurement with probability $1-p$. Eventually, we take the limit $\Delta t\to 0$, and average over all measurement outcomes. The parameter $\gamma$ thus plays the role of a measurement rate. As we show in Appendix~\ref{App.Master}, this dynamics is described by the Lindblad equation
\begin{equation}
\partial_t \hat{\rho}(t)= -i[\hat{H},\hat{\rho}(t)] +\gamma \left( \frac{1}{2\pi} \int_0^{2\pi} d\alpha e^{i\alpha \hat{Q}_A} \hat{\rho}  e^{-i\alpha \hat{Q}_A} -\hat{\rho} \right)\,,
\label{eq:Lindblad}
\end{equation}
which is valid for arbitrary choices of the interval $A$. We stress that in this work, rather than considering single monitored trajectories, we focus on the dynamics of the resulting average mixed state. As shown in Appendix \ref{App.Master}, the latter is completely positive and trace-preserving (CPTP).
In the following, we will focus on the case where $A$ coincides with half of the system, say, the right one. This choice is natural as it respects the geometry of the bipartition protocol, but we will also comment on how the dynamics is modified if $A$ is chosen differently. We note that a similar bipartite setting was studied in Ref.~\cite{AlbaCarollo2022LocalizedLosses}, which however considered a localized (single site) loss process. As we will see, our work differs from Ref.~\cite{AlbaCarollo2022LocalizedLosses} also by the fact that we are interested in describing the system at large space-time scales, rather than in finding an exact solution to the microscopic dynamics.

Note that measuring $\hat Q_A$ is not equivalent to measuring $\hat n_j$ over all $j\in A$. For instance, when $A$ coincides with the whole system, $\hat Q_A=\hat{Q}$ is conserved. Then, if we initialize the system in a state with a well-defined number of particles, measuring $\hat Q_A$ does not affect the dynamics. Conversely, measuring $\hat n_j$ over all $j$ resets the system to a Fock state in real space. Therefore, our work differs from previous studies focusing on the monitoring of the particle number $\hat{n}_j$ over extensive regions, see e.g.~\cite{cao2019entanglement,alberton2021entanglement,fava2023nonlinear,poboiko2023theory,fava2024monitored}. In fact, it is useful to stress that a projective measure of $\hat Q_A$ does not preserve the Gaussianity of the system~\cite{bravyi2004lagrangian}, making it difficult to investigate individual quantum trajectories.

While the Hamiltonian is quadratic in the fermions, the Lindbladian~\eqref{eq:Lindblad} does not preserve the Gaussianity of the density matrix. However, its structure is such that local correlation functions satisfy closed hierarchies of
equations of motion, see Refs.~\cite{eisler2011crossover,Zunkovic2014ClosedHierarchy,caspar2016dissipative,Caspar2016dynamics,Hebenstreit2017Solvable,Foss-Feig2017solvable,klich2019closed} for a detailed discussion on the conditions under which this simplification occurs. In our case, single-body correlation functions are trivial, so that we will focus on the dynamics of the two-body correlation functions
\begin{equation}
    C_{n,m}(t)={\rm Tr}[\hat{\rho}(t) \hat{c}^\dagger_n \hat{c}_m]=:\langle\hat{c}^\dagger_n \hat{c}_m\rangle_t\,.\label{eq:cov_mat_def}
\end{equation}
Using Eq.~\eqref{eq:Lindblad}, a standard derivation yields
\begin{equation}
\label{eq:ODE_C_ij_explicit}
\frac{dC_{i,j}}{dt} = i J (C_{i,j+1} + C_{i,j-1} - C_{i-1,j} - C_{i+1,j}) -\gamma C_{i,j} + \gamma g_A(i,j) C_{i,j}\,,
 \end{equation}
where $g_A(i,j)=1$ if $i,j \in A$ or $i,j \in \bar{A}$ and $g_A(i,j)=0$ otherwise.
Denoting by $C(t)$ the matrix whose elements are defined in Eq.~\eqref{eq:cov_mat_def}, we can also rewrite the system in matrix form as
\begin{equation}
    \partial_t C=  i[H, C] -\gamma (P_A CP_{\bar{A}} + P_{\bar{A}}CP_A)\,, \label{eq:EOM_C}
\end{equation}
where
$(P_A)_{m,n} = \delta_{m,n}$ if $m,n \in A$ and $(P_A)_{m,n}=0$ otherwise, while $P_{\bar{A}}=1-P_A$. 

Although the system of equations~\eqref{eq:ODE_C_ij_explicit} is not integrable, the computational cost of its numerical solution scales only polynomially, rather than exponentially, in the system size. This is a major simplification compared to typical Lindbladian dynamics, allowing us to numerically obtain $C(t)$ up to large values of $N$ and $t$ by means of standard numerical methods. In the next sections we will present explicit numerical data obtained by solving Eqs.~\eqref{eq:ODE_C_ij_explicit} and use them to study the dynamics of bipartition protocols under extensive-charge measurements.

The qualitative description of our main results is the following. Starting from bipartite product states, we observe the emergence of a ballistic regime for $x,t \to \infty$, where the profiles of local observables are described in terms of the ray variable $\zeta= x/t$. In contrast to the unitary case, extensive charge monitoring induces discontinuities at $\zeta=0$ in the average state. Remarkably, these discontinuities are observed also when the initial state is a homogeneous thermal state. We develop a general theory to understand this phenomenology in terms of ballistic GHD with modified boundaries. Our solution is based on a hybrid analytical-numerical method, where the microscopic information required to fully fix the behavior of the local observables around the discontinuity is provided by a numerical analysis of the continuity equations around the point $x=0$.
\section{The GHD framework}
\label{sec:GHD}

In this section, we introduce a GHD framework to analyze the dynamics of the system at large space-time scales. After recalling the standard GHD approach to bipartitioning protocols under unitary dynamics, we extend it to the case of extensive-charge monitoring.

\subsection{The unitary bipartition protocol}

We begin by recalling the main features of the unitary bipartition protocol and its GHD solution in integrable systems. We will only recall the aspects that are directly relevant for our work (and specialized to free fermions), referring to the literature for a more comprehensive discussion (see in particular the review~\cite{alba2021generalized} and references therein).

After initializing the system in the bipartite state~\eqref{eq:initial_state}, an integrable system is characterized by the emergence of a ballistic light-cone propagating at finite velocity from the junction. After a transient time, the system locally equilibrates to space-time dependent quasi-stationary states described by a generalized Gibbs ensemble (GGE)~\cite{VidmarRigol2016GGEReview} or, equivalently, by its quasi-particle momentum distribution functions $n_{x,t}(k)$. The GHD allows us to compute such functions and obtain a prediction for the space-time profiles of all local observables.

In essence, the GHD approach starts from the set of continuity equations for the infinitely-many conserved charges
\begin{equation}\label{eq:discrete_continuity_free}
    \partial_t \langle \hat q_x^{(r,\pm)}\rangle_t=\langle \hat J^{(r,\pm)}_{x-1}\rangle_t-\langle \hat J^{(r,\pm)}_{x}\rangle_t\,,
\end{equation}
where
\begin{align}
\hat  J^{(r,+)}_x &= i J^2\left[ \left(\hat c^\dagger_{x+1}\hat c_{x+r}-\hat c^\dagger_{x}\hat c_{x+r+1}\right)+ (\hat c^\dagger_{x+r+1}\hat c_{x}-\hat c^\dagger_{x+r}\hat c_{x+1})\right]\,, \label{eq:jp}\\
     \hat J^{(r,-)}_x &= - J^2\left[\left(\hat c^\dagger_{x+1}c_{x+r}-\hat c^\dagger_{x}\hat c_{x+r+1}\right)- (\hat c^\dagger_{x+r+1}\hat c_{x}-\hat c^\dagger_{x+r}\hat c_{x+1})\right]\,,\label{eq:jm}
\end{align}
are the current densities associated with the local charges.
Next, recall that, given an arbitrary function $n(k)$ corresponding to a GGE, the expectation values of the local densities and currents can be written as~\cite{alba2021generalized}
\begin{align}
    \langle \hat{q}^{(r,\pm)}\rangle_{n}&=\int^{+\pi}_{-\pi} dk \, q^{r,\pm}(k) n(k), \label{eq:ghd_q}\\
    \langle \hat{J}^{(r,\pm)}\rangle_{n} &=\int^{+\pi}_{-\pi} dk \, \varepsilon'(k) q^{r,\pm}(k) n(k)\,,\label{eq:ghd_j}
\end{align}
where $\varepsilon(k)$ is the single-particle energy~\eqref{eq:H_esp}. Notice that, choosing the Hamiltonian as in Eq. \eqref{eq:Hamiltonian}, $\varepsilon'(k)=2J\sin(k)$. We now assume that each mesoscopic region (or ``fluid cell'') traveling at a given velocity $v$ from the junction, equilibrates for $t\to\infty$ to a $v$-dependent GGE~\cite{alba2021generalized}. Then, plugging Eqs.~\eqref{eq:ghd_q} and~\eqref{eq:ghd_j} into Eq.~\eqref{eq:discrete_continuity_free}, and introducing the rescaled variable $\zeta=x/t$, the continuity equations become
\begin{equation}\label{eq:charge_continuity_free}
    -\zeta \partial_\zeta \langle \hat q^{(r,\pm)}\rangle_{n_\zeta} +\partial_{\zeta}\langle \hat J^{(r,\pm)}\rangle_{n_\zeta}=0\,.
\end{equation}
Since Eq.~\eqref{eq:charge_continuity_free} holds for all charges, and the charges are complete, we arrive at the GHD equation
\begin{equation}\label{eq:GHD_k}
    (-\zeta+\varepsilon^\prime(k))\partial_\zeta n_\zeta(k)=0\,.
\end{equation}

For a fixed $k$, the solution to Eq.~\eqref{eq:GHD_k} is a piece-wise constant function, with a jump occurring at $\zeta=\varepsilon^\prime(k)$. Crucially, the GHD equation~\eqref{eq:GHD_k} must be supplemented with the boundary condition
\begin{equation}\label{eq:boundary_cond}
    \lim_{\zeta\to + \infty} n_\zeta(k)=n^R(k)\,, \quad \lim_{\zeta\to - \infty} n_\zeta(k)=n^L(k)\,,
\end{equation}
where $n^{L/R}(k)$ are the quasi-particle momentum distribution functions corresponding to the GGEs associated with the left and right initial states, respectively. Eq.~\eqref{eq:boundary_cond} encodes the physical requirements that, infinitely far right (left) from the junction, the system relaxes to the GGE corresponding to a homogeneous quench from the initial state $\hat\rho_R$ ($\hat\rho_L)$. Putting all together, one arrives at the GHD solution
\begin{equation}\label{eq:ghd_solution_free}
    n_\zeta(k)=\Theta(\varepsilon^\prime(k)-\zeta) n^{L}(k)+\Theta(-\varepsilon^\prime(k)+\zeta) n^{R}(k)\,,
\end{equation}
where $\Theta(x)$ is the Heaviside theta function. Physical predictions for the profiles of local charges and currents can then be obtained from Eqs.~\eqref{eq:ghd_q}, ~\eqref{eq:ghd_j}, and similarly for more general local observables. 

\subsection{The bipartition protocol under extensive-charge monitoring}\label{sec:ghd_monitored_dyn_theory}

We now present our first main result, namely we show that the GHD framework can be extended to describe the monitored dynamics introduced in Sec.~\ref{sec:the_protocol}. This observation is a priori non-trivial, since the charge measurements generate non-local, integrability-breaking terms in the equations for the two-point correlation functions~\eqref{eq:ODE_C_ij_explicit}. 

As already mentioned, the Lindbladian appearing in Eq.~\eqref{eq:Lindblad} does not display any non-trivial local conservation law. Nevertheless, we may study how the densities of the Hamiltonian charges evolve in time. From Eq.~\eqref{eq:Lindblad}, an elementary calculation yields 
\begin{equation}\label{eq:discrete_continuity}
    \partial_t \langle \hat q_x^{(r,\pm)} \rangle= \langle \hat J^{(r,\pm)}_{x-1} \rangle -\langle 
    \hat J^{(r,\pm)}_{x}\rangle -\gamma \Pi^{(r)}_x \langle \hat q_x^{(r,\pm)}\rangle \,,
\end{equation}
where $\hat J^{(r,\pm)}_{x}$ are defined in Eqs.~\eqref{eq:jp}, and ~\eqref{eq:jm}, while 
\begin{equation}
\label{eq:projector_Lorenzo}
    \Pi^{(r)}_x=
    \begin{cases}
        1 & {\rm if\ \ }  x\in\{-r+1,\ldots 0\}\,, \\
        0 & {\rm otherwise.}
    \end{cases}
\end{equation}
Note that the term proportional to $\gamma$ is absent for $r=0$. Eq.~\eqref{eq:discrete_continuity} tells us that local charges satisfy the standard continuity equation governing the unitary dynamics, up to a finite number of sink-like terms stemming  from the incoherent charge measurement. Notice also that these terms are absent in the continuity equations for the charges $\hat q_{x}^{(0,+)}$. It is therefore natural to assume that, away from the junction, one should recover a picture in terms of local quasi-stationary states described by a GGE.

As in the unitary case, we then assume that, for $\zeta\neq 0$, each fluid cell traveling at velocity $\zeta$ from the junction equilibrates to a $\zeta$-dependent GGE. Accordingly, Eq.~\eqref{eq:discrete_continuity} yields
\begin{equation}\label{eq:charge_continuity}
    -\zeta \partial_\zeta \langle \hat q^{(r,\pm)}\rangle_{n_\zeta} +\partial_{\zeta}\langle \hat J^{(r,\pm)}\rangle_{n_\zeta}=0\,,   \qquad (\zeta\neq 0)\,.
\end{equation}
Namely, repeating the argument of the previous section, and separating the cases of positive and negative values of $\zeta$, we have the set of GHD equations
\begin{align}
        (-\zeta+\varepsilon^\prime(k))\partial_\zeta n_\zeta(k)&=0\qquad \zeta>0\,,\label{eq:one}\\
        (-\zeta+\varepsilon^\prime(k))\partial_\zeta n_\zeta(k)&=0\qquad \zeta<0\,.\label{eq:two}
\end{align}

For a given $k$, the solution to Eqs.~\eqref{eq:one}, \eqref{eq:two} is again piece-wise continuous. However, an additional discontinuity appears at $\zeta=0$. Quantitatively, fixing $k>0$, it follows that $(-\zeta+\varepsilon^\prime(k))>0$ for $\zeta<0$ and so Eq.~\eqref{eq:two} implies $\partial_\zeta n_\zeta(k)=0$ for $\zeta<0$ (note that, due to the choice of the Hamiltonian, we have ${\rm sign} [\varepsilon^\prime(k)]={\rm sign} [\sin(k)]={\rm sign}[k]$). Therefore, $n_\zeta(k)$ is a constant for $\zeta<0$ and, as argued in the previous section, must be equal to $n^L(k)$, cf. Eq.~\eqref{eq:boundary_cond}. Next, for $\zeta>0$, Eq.~\eqref{eq:one} implies that $n_\zeta(k)$ is piece-wise constant, with a possible discontinuity at $\zeta^\ast=\varepsilon^\prime(k)$. Again, for $\zeta>\zeta^\ast$ the constant is given by $n^R(k)$, while for $0<\zeta<\zeta^\ast$ we have an unknown value $\chi^R(k)$ (not depending on $\zeta$). Proceeding similarly for $k<0$, we arrive at the formal GHD solution
\begin{empheq}{equation}\label{eq:sol}
n_\zeta(k)= 
\begin{cases}
 \Theta(-\zeta) n^L(k)+ \Theta(\zeta)\Theta(\varepsilon^\prime(k)-\zeta) \chi^R(k)+ \Theta(\zeta-\varepsilon^\prime(k))n^R(k)  & k>0\,,\\
 \Theta(\zeta) n^R(k)+ \Theta(-\zeta)\Theta(-\varepsilon^\prime(k)+\zeta) \chi^L(k)+ \Theta(\varepsilon^\prime(k)-\zeta)n^L(k) & k<0\,,
   \end{cases}
\end{empheq}
which depends on the unknown function
\begin{equation}\label{eq:chi_function}
    \chi(k)=
    \begin{cases}
        \chi^R(k)\,, & k\geq 0\,,\\
        \chi^L(k)\,, & k< 0\,.
    \end{cases}
\end{equation}

Eq.~\eqref{eq:sol} has a very natural interpretation in terms of the so-called quasi-particle picture~\cite{calabrese2005evolution, alba2021generalized}. Let us focus, for instance, on $k>0$, $\zeta> 0$. Eq.~\eqref{eq:sol} states that the quasi-particle content $n_\zeta(k)$ of the GGE coincides with $n^R(k)$ if $\zeta$ is greater than the quasi-particle velocity $\varepsilon'(k)$ (no quasi-particle coming from the junction can contribute). Conversely, if $\zeta\leq \varepsilon'(k)$ the rapidity distribution function is modified in a non-trivial way, since particles coming from the junction and undergoing a non-trivial evolution contribute. In fact, Eq.~\eqref{eq:sol} has the same form encountered in the study of bipartition protocols with an additional Hamiltonian impurity, or defect, localized at the junction~\cite{bastianello2018nonequilibrium,LjubotinaZnidaricProsen2019Defect,delVecchio2022transport,gouraud2022quench,gouraud2023stationary, capizzi2023domain,rylands2023transport,capizzi2023entanglement}. Intuitively, this similarity may have been expected, given the form of the continuity equations~\eqref{eq:discrete_continuity}. We note, however, that this is not a perfect analogy as the dynamics studied here is non-unitary. 

In the context of Hamiltonian impurities, the unknown function $\chi(k)$ can be derived by computing the transmission and reflection coefficients in the single-particle sector, via a scattering-matrix approach~\cite{LjubotinaZnidaricProsen2019Defect,rylands2023transport}, see also~\cite{bertini2016determination}. This approach, however, appears problematic in our setting. The difficulty is mainly due to the fact that the two-body Lindbladian appearing in Eq.~\eqref{eq:Lindblad} is non-local and cannot be diagonalized analytically. In addition, the non-unitarity of the evolution makes it not clear how to adapt the standard scattering-matrix approach used in previous studies to our setting. For these reasons, we have followed a different approach targeting directly the unknown function $\chi(k)$. The idea is to derive a set of merging conditions for the hydrodynamic quasi-particle distribution functions $n_\zeta(k)$ at $\zeta=0^{\pm}$. In turn, such merging conditions provide a consistency equation for $\chi(k)$, which admits a unique solution.

To be precise, consider an interval $I_t$ around the origin with end points given by $\pm \nu t$. Summing  Eq.~\eqref{eq:discrete_continuity} over $j\in I_t$ yields
\begin{equation}\label{eq:partial_cont_eq}
    \sum_{x=-\nu t}^{\nu t} \partial_t \langle \hat{q}^{(r,\pm)}_x(t)\rangle= \langle \hat{J}^{(r,\pm)}_{-\nu t-   1 }\rangle- \langle \hat{J}^{(r,\pm)}_{\nu t}\rangle  - \gamma \sum_{x=-r+1}^0 \langle \hat{q}^{(r,\pm)}_x(t) \rangle\,,
\end{equation}
where we assumed that time is large enough, so that $\nu t > r$. Next, setting $\zeta= x/t$, to the leading order in $1/t$ we can approximate
\begin{equation}
    \sum_{x=-\nu t}^{\nu t} \langle \hat{q}^{(r,\pm)}_x(t)\rangle\simeq t \int_{-\nu}^\nu d\zeta    \langle \hat{q}^{(r,\pm)}\rangle_\zeta\,.
\end{equation}
Therefore, taking the limit $\nu \to 0$, we obtain
\begin{equation}
    \lim_{\nu \to 0}    \sum_{x=-\nu t}^{\nu t} \partial_t \langle \hat{q}^{(r,\pm)}_x(t)\rangle= \lim_{\nu \to 0} \int_{-\nu}^\nu d\zeta    \langle \hat{q}^{(r,\pm)}\rangle_\zeta=0\,.
\end{equation}
Finally, plugging this equation into Eq.~\eqref{eq:partial_cont_eq},  we arrive at the merging conditions
\begin{equation}\label{eq:mergin_equations}
\langle \hat{J}^{(r,\pm)}   \rangle_{0^+} - \langle \hat{J}^{(r,\pm)}\rangle_{0^- }= -\gamma \mathcal{Q}(r,\pm)\,,
\end{equation}
where, we define
\begin{equation}
   \mathcal{Q}(r,\pm) := \sum_{x=-r+1}^0 \langle \hat{q}^{(r,\pm)}_x(\infty) \rangle . \label{eq:Q_bdy}
\end{equation}

The merging conditions feature the quantity $\mathcal{Q}(r,\pm)$ which, in turn, depends on the initial state in a non-trivial way. Unfortunately, computing the value of $\mathcal{Q}(r,\pm)$ is hard, representing a bottle-neck to arrive at a fully analytic solution of the quench protocol. Nevertheless, the quantities $\mathcal{Q}(r,\pm)$ can be computed numerically solving the system of equations~\eqref{eq:ODE_C_ij_explicit} up to late times. The numerical values computed in this way can be plugged into Eq.~\eqref{eq:mergin_equations}, yielding an explicit set of merging conditions that determine the function $\chi(k)$.

Before proceeding, a few comments are in order. First, we emphasize that the approach described above is hybrid, requiring to first extract some microscopic data by numerical computations (the values $\mathcal{Q}(r,\pm)$), and then to perform analytic calculations to solve the GHD equations under the the merging conditions~\eqref{eq:mergin_equations}. It is important to stress this method provides more information compared to a brute-force numerical solution to the microscopic dynamics. Indeed, the solution to Eqs.~\eqref{eq:ODE_C_ij_explicit} only gives us access to the profiles of observables expressed in terms of two-body correlation functions. Instead, our approach yields the densities $n_\zeta(k)$, which completely describes the state of the system in the fluid cell traveling at speed $\zeta$.

Second, our GHD approach above gives us access to the \emph{exact} profiles in the hydrodynamic limit $t,N\to\infty$. In this respect, we note that obtaining an accurate numerical approximation for $\mathcal{Q}(r,\pm)$ requires an amount of computational resources which is less than the one required to obtain an accurate prediction for the full profiles of arbitrary local observables. This will be apparent from our numerical results presented in Sec.~\ref{sec:sub_numerics}. Our data will generally show that at the time scales at which the profiles approach their stationary values near $\zeta=0$ (so that $\mathcal{Q}(r,\pm)$ can be extracted in a reliable way), the profiles still show large finite-time effects at large distances from the origin.

We also note that, in the actual implementation of the above method, one needs to introduce a cut-off $r_{\rm cut}$, because $\mathcal{Q}(r,\pm)$ can only be computed for a finite number of values $r=1,\ldots r_{\rm cut}$. Then, one obtains an approximate solution $\chi^{(r_{\rm cut})}(k)$ by setting $\mathcal{Q}(r,\pm)=0$ for $r>r_{\rm cut}$. As we will see, for the quench protocols that we will consider, the asymptotic expectation values of higher local charges $\langle \hat q_{x}^{r,\pm}(\infty)\rangle $ quickly vanish as $r$ increases, so that the accuracy of $\chi^{(r_{\rm cut})}(k)$ is expected to be very high already for relatively small $r_{\rm cut}$.

Finally, we highlight that, while this discussion is completely general, in some cases one may exploit additional symmetries or properties of the initial states to obtain an exact or very accurate ansatz for $\mathcal{Q}(r,\pm)$. Alternatively, one may be able rewrite the merging conditions in such a way that the contributions from the microscopic data is small. We will follow this strategy for domain-wall initial states. In this case, we will be able to conjecture a reformulation of the merging conditions which turns out to be very accurate even without any input from microscopic data (namely, for a cut-off value $r_{\rm cut }=0$).

Before leaving this section, we mention that a similar GHD framework can be adapted for different choices of the region $A$. We found in particular a qualitatively similar phenomenology when the interval $A$ consists of a finite number of sites (even a single one) distributed symmetrically near the junction.

\section{GHD solution from domain-wall initial states}
\label{sec:micro_general}

In this section we present a  solution to the monitored dynamics starting from special families of bipartite pure initial states.  We focus on states of the form
\begin{equation}\label{eq:initial_state_DW}
    \ket{\Psi_0}=\ket{0}_L\otimes \ket{\psi_0}_R\,,
\end{equation}
where $\ket{0}_L$ is the vacuum state, while $\ket{\psi_0}_R$ is a non-trivial product state. For concreteness, we will focus on the so-called N\'eel state
\begin{equation}
\ket{\psi^N_0}_R=\prod_{j=0}^{\lfloor N/2 \rfloor-1} \hat{c}^\dagger_{2j+1} |0\rangle\,.\label{eq:initial_vacuum-neel}
\end{equation}
However, as we will specify later, the calculations presented here hold for more general initial states, including the fully polarized state 
\begin{equation}\label{eq:fully_polarized}
    \ket{\psi^P_0}_R=\prod_{j=1}^{(N-1)}\hat{c}^\dagger_{j} |0\rangle\,.
\end{equation}
We note that the dynamics from such bipartite initial states have been extensively studied in the past, both for isolated and open quantum systems~\cite{alba2021generalized,Alba2025DephasingBoundaryDriving, AlbaCarollo2022LocalizedLosses, AntalKrapivskyRakos2008LogCurrentFluct, Schoenhammer2007FCSLevitovLesovik, Schoenhammer2009FCSFiniteT, medvedyeva2016Exact,
eisler2011crossover, TurkeshiSchiro2021BoundaryDephasing,
LjubotinaZnidaricProsen2019Defect,DolgirevMarinoSelsDemler2020NonGaussianDephasing,
DiFrescoEtAl2024DarkIntervals,RussottoAresCalabreseAlbaQSSEP,scopa2023scaling}

\subsection{The numerical solution}\label{sec:sub_numerics}

Before illustrating our GHD approach, we present our numerical solution to the Lindblad equation~\eqref{eq:ODE_C_ij_explicit}. The profiles obtained in this way confirm the assumptions stated in the previous section, underlying the validity of the GHD framework. In addition, the numerical data will serve to test quantitatively our analytic predictions. 

\begin{figure}
    \centering
\includegraphics[height=0.19\textheight,width=1.006\textwidth]{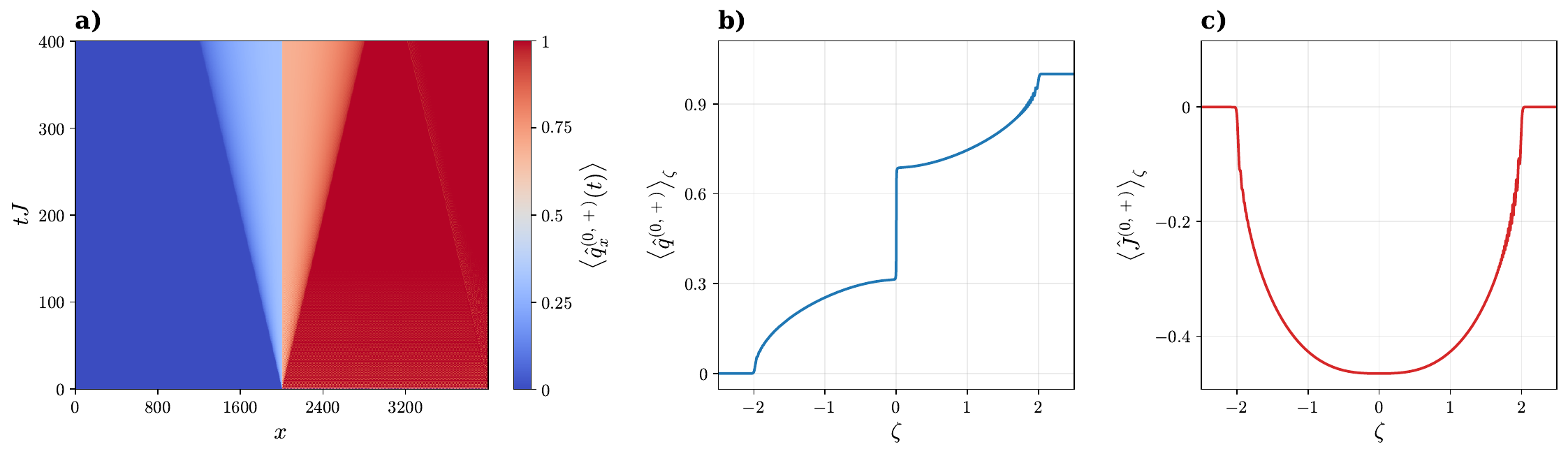}
    \caption{a) Non-unitary dynamics of the charge $\langle \hat{q}_x^{(0,+) }(t)\rangle$. The system is prepared in a domain-wall state given by Eq.~\eqref{eq:initial_state_DW}. A light cone due to ballistic propagation of quasi-particles is present with front at $\zeta^*= \pm {\rm max}(|v(k)|)=\pm 2J$.
    b) Late time profile of the local charge $\langle \hat{q}_x^{(0,+) }(t)\rangle$ for $\zeta=x/t$. Far from the propagation front ray $\zeta^*$, the charge profile remains constant, while it displays a discontinuity at $\zeta=0$.  c) Late time profile of the local current $\langle \hat{J}_x^{(0,+) }(t)\rangle$. The current is continuous for all ray values.  We set $J=1$, $\gamma=1$, $L=2N= 4000$ sites,  and $tJ=400$ for calculating the charge and current profiles. }
    \label{fig:r0_profiles_vacuum-neel}
\end{figure}

First, note that the initial conditions of Eqs.~\eqref{eq:ODE_C_ij_explicit} are easily obtained from the initial state. In particular, focusing on the bipartite state defined by Eqs.~\eqref{eq:initial_state_DW} and~\eqref{eq:initial_vacuum-neel}, the initial condition corresponds to
\begin{equation}
C_{x,y}(0)=\begin{cases}\label{eq:initial_condition}
0, &\quad x,y \le 0, \\
\delta_{x,y}, &\quad x,y >0\,,\ x\equiv 0\ \ ({\rm mod}\ 2)\,.
\end{cases}
\end{equation}
As mentioned, the system~\eqref{eq:ODE_C_ij_explicit} with initial conditions~\eqref{eq:initial_condition} can be solved via standard numerical methods. We have done so up system sizes $2N\sim 10000$ and times $t\sim 2000$, for different values of $\gamma$. Given the solution for $C_{x,y }(t)$, it is immediate to obtain the profiles of local charges and currents, as they are quadratic in the fermions. An example of our data is displayed in Fig.~\ref{fig:r0_profiles_vacuum-neel}.

In the plots, we rescale the profiles introducing the variable $\zeta=x/t$. We have verified an excellent data collapse as $t$ increases, fully confirming the GHD assumptions for any value of $\gamma$. As a general feature, the plots show a discontinuity at $\zeta=0$, which becomes more pronounced as $\gamma$ increases. 
We note that the initial particle imbalance induced by the domain-wall condition is responsible for transport of quasi-particles from the right-half towards the left one. The presence of the discontinuity in the charge profile indicates that at the junction, extensive monitoring of the subsystem charge induces non-trivial quasi-particle scattering.
Additional plots are provided in Sec.~\ref{sec:sub_ghd_analytic}.

Numerically, we observe that the profiles of some charges are identically vanishing. This observation can be explained by symmetry considerations. To do so, we first introduce the unitary and anti-unitary operators $\hat{U}$ and $K$, respectively defined by $\hat{U}= e^{i\pi \sum_l  l\hat{n}_l}$ and $K^{\dagger}i K=-i$, where $i$ is the imaginary unit. Then, it is easy to show that $\hat \Theta= \hat{U}K$ is a symmetry of the Lindbladian. Indeed, since $\hat U\hat{c}_j \hat U^{-1}= (-1)^j \hat{c}_j$ and $\hat U\hat{n}_j \hat U^{-1}=  \hat{n}_j $, we obtain
\begin{align}
    \hat U \hat{H}\hat U^{-1}&= -\hat{H} ,\\
   \hat U\hat{Q}_A\hat U^{-1}&= \hat{Q}_A .
\end{align}
Plugging this into the Lindbladian, we find that
\begin{align}
    \hat{\Theta }\mathcal{L}(\hat{\rho})\hat{\Theta}^{-1}&= -i[\hat{H}, \hat{\Theta}\hat{\rho}\hat{\Theta}^{-1}] -\gamma \left\{ \int_{-\pi}^\pi \frac{d\alpha}{2\pi} e^{i\alpha \hat{{Q}_A}}\hat{\Theta}\hat{\rho}\hat{\Theta}^{-1}e^{-i\alpha \hat{{Q}_A}} -\hat{\Theta }\hat{\rho}\hat{\Theta}^{-1}\right\}\nonumber\\
    &= \mathcal{L}(\hat{\Theta }\hat{\rho}\hat{\Theta}^{-1})\,.\label{eq:weak_symm}
\end{align}
Since the pure initial state $\hat\rho_0 = \ketbra{\Psi_0}{\Psi_0}$, with $\ket{\Psi_0}$ defined by Eqs.~\eqref{eq:initial_state_DW} and~\eqref{eq:initial_vacuum-neel} is an eigenstate of $\hat U$, we have
\begin{equation}
      \hat{\Theta }\hat{\rho}_t\hat{\Theta}^{-1}=\hat{\rho}_t\,,
\end{equation}
implying that for every observable $\hat{\mathcal{O}}$, $\langle \hat{\Theta }\hat{\mathcal{O}}\hat{\Theta}^{-1}\rangle_t=\langle\hat{\mathcal{O}}\rangle_t$. On the other hand, from the explicit expressions Eqs. \eqref{eq:qp}, \eqref{eq:qm}, we obtain
\begin{align}
   \langle \hat{\Theta} \hat{q}^{(2r,-)}_x\hat{\Theta}^{-1} \rangle_t &=-\langle \hat{q}^{(2r,-)}_x \rangle_t \,,\\
    \langle \hat{\Theta} \hat{q}^{(2r+1,+)}_x\hat{\Theta}^{-1} \rangle_t &=-\langle \hat{q}^{(2r+1,+)}_x \rangle_t\,,
\end{align}
hence
\begin{equation}\label{eq:vanishing_condition}
\langle \hat{q}^{(2r,-)}_x \rangle_t = \langle \hat{q}^{(2r+1,+)}_x \rangle_t=0\,.
\end{equation}

The rest of this section is devoted to provide a prediction for the non-zero profiles, following the GHD approach developed in Sec.~\ref{sec:ghd_monitored_dyn_theory}. In detail, after deriving an equivalent formulation of the merging conditions~\eqref{eq:mergin_equations} in Sec.~\ref{sec:sub_merging}, we will present the full GHD solution in Sec.~\ref{sec:sub_ghd_analytic}.

\subsection{The merging conditions}\label{sec:sub_merging}

\begin{figure}
    \centering
    \includegraphics[width=\linewidth]{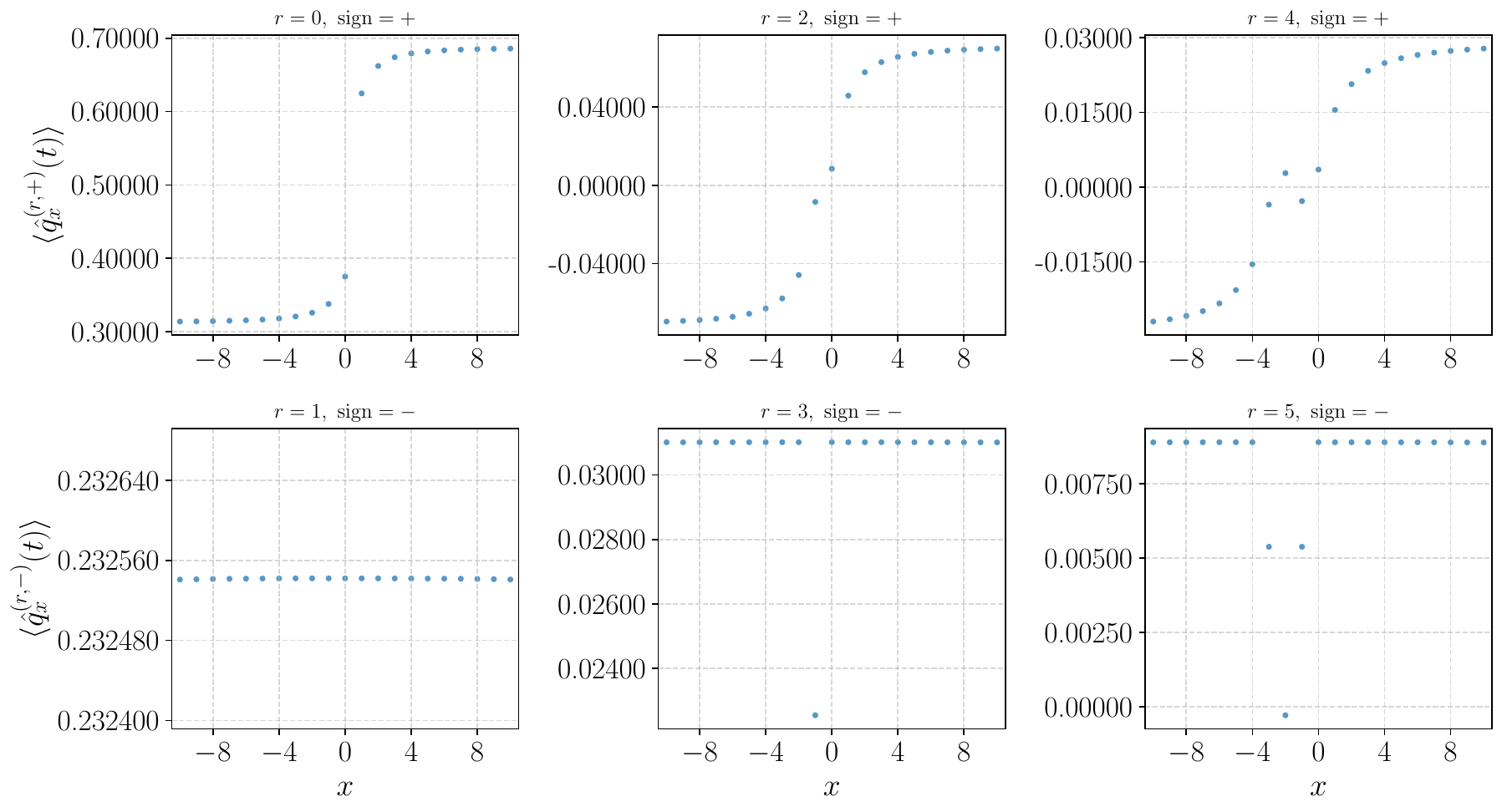}
    \caption{Values of the local charges in the NESS. The plots only show the charges that are not identically zero. The data correspond to $J=\gamma=1$ and $N=4000$, while time is taken to be $tJ=400$, which is large enough that the data points appear to be converged up to a precision of at least $10^{-4}$.}
    \label{fig:NESS_profiles}
\end{figure}

As discussed in Sec.~\ref{sec:ghd_monitored_dyn_theory}, the first step of our GHD approach is to compute the values $\mathcal{Q}(r,\pm)$ appearing in Eq.~\eqref{eq:mergin_equations}, to obtain a valid set of merging conditions. 

In order to find some possible hidden structure, we first study numerically the behavior of $\langle \hat{q}_x^{(r,\pm))}(\infty)\rangle$ near the origin, $x=0$. Fig.~\ref{fig:NESS_profiles} reports an example of our data, from which we see that the profiles $\langle \hat{q}_x^{(r,-)}(\infty)\rangle$ are approximately constant as a function of $x$, and in any case appear to be constant away from a finite number of points localized near the origin. This observation suggests that, for the initial state under consideration, we may be able to reformulate the merging conditions in a way that makes our approach simpler, as we explain below.

Indeed, suppose first that the profiles $\langle \hat q_x^{(r,-)}(\infty)\rangle$ are exactly constant as a function of $x$. The set of expectation values $\{\langle \hat q_x^{(r,\pm )}(\infty)\rangle\}_{x\in \mathbb{Z}}$ characterize the non-equilibrium steady state (NESS) associated with the bipartition protocol, and are related to the hydrodynamic profiles via the identification~\cite{bertini2016determination}
\begin{equation}
    \lim_{x \to\pm\infty} \langle \hat q_x^{(r,\pm)}(\infty)\rangle=\langle  \hat q^{(r,\pm)}\rangle_{\zeta=0^\pm} \label{eq:zeta0lim}
\end{equation}
Therefore, the fact that $\langle \hat q_x^{(r,-)}(\infty)\rangle$ are constant implies
\begin{equation}
        \langle \hat{q}^{(r,-)}\rangle_{\zeta={0^-}} = \langle \hat{q}^{(r,-)} \rangle_{\zeta={0^+}}=:\langle \hat{q}^{(r,-)} \rangle_{\zeta={0}}\,.
        \label{eq:cond1} 
\end{equation}
Under the same assumptions, Eq.~\eqref{eq:mergin_equations} becomes
\begin{align}
    \langle \hat{J}^{(r,-)} \rangle_{\zeta={0^+}} -\langle \hat{J}^{(r,-)} \rangle_{\zeta={0^-}}&=-\gamma r \langle \hat{q}^{(r,-)} \rangle_{0}. \label{eq:cond2_ideal}
\end{align}

From inspection of the numerical data, we found that the condition
\begin{equation}
        \lim_{x\to -\infty} \langle \hat q_x^{(r,-)}(\infty)\rangle=         \lim_{x\to +\infty} \langle \hat q_x^{(r,-)}(\infty)\rangle\,,
\end{equation}
appears to be true up to numerical precision, so that Eq.~\eqref{eq:cond1} is indeed verified for our initial state. Unfortunately, however, it is apparent from the data in Fig.~\ref{fig:NESS_profiles} that the profiles $\langle \hat q_x^{(r,-)}(\infty)\rangle$ are not constant near the origin.
This discrepancy is not obvious from the first charge profile, but it becomes more pronounced as we increase the parameter $r$, controlling the locality degree of the conserved charges. Therefore, $\mathcal{Q}(r,-)\neq r \langle \hat q_x^{(r,-)}(\infty)\rangle$, and the merging condition~\eqref{eq:cond2_ideal} should be replaced by
\begin{align}
    \langle \hat{J}^{(r,-)} \rangle_{\zeta={0^+}} -\langle \hat{J}^{(r,-)} \rangle_{\zeta={0^-}}&=-\gamma r \langle \hat{q}^{(r,-)} \rangle_{0}+\delta_r\,. \label{eq:cond2}
\end{align}
The numbers $\delta_r$ takes into account the deviations of $\langle \hat q_x^{(r,-)}(\infty)\rangle$ from being a constant in the vicinity of $x=0$.

Eqs.~\eqref{eq:cond1}, \eqref{eq:cond2} represent the correct merging condition for our problem. As we will show in the next section, they uniquely fix the unknown function $\chi(k)$ in Eq. \eqref{eq:chi_function} and can therefore be considered an equivalent reformulation of Eqs.~\eqref{eq:mergin_equations}.

As mentioned in Sec. \ref{sec:ghd_monitored_dyn_theory}, we will compute numerically the numbers $\{\delta_r\}_{r=1}^{r_{\rm cut}}$, where $r_{\rm cut}$ is a cut-off, and set $\delta_r=0$ for $r> r_{\rm cut}$. We will then obtain a function $\chi^{(r_{\rm cut})}(k)$ for the GHD solution depending on $r_{\rm cut}$. Remarkably, we will show that $\chi^{(0)}(k)$ is already an excellent approximation to the exact solution, and $\chi^{(r_{\rm cut})}(k)$ quickly converges as $r_{\rm cut}$ is increased. This observation motivates the reformulation of the merging conditions carried out above. We refer to Appendix~\ref{sec:numeric_micro_extraction} for details on how the numbers $\delta_r$ are extracted numerically. 

\subsection{The GHD solution}\label{sec:sub_ghd_analytic}

In this section, we finally provide an analytic solution for the monitored dynamics. Following the general strategy outlined in Sec.~\ref{sec:ghd_monitored_dyn_theory}, we determine the unknown function $\chi(k)$, appearing in the formal solution~\eqref{eq:sol}, using the merging conditions~\eqref{eq:cond1} and~\eqref{eq:cond2}. Below, we only present the main steps of the derivation, referring to Appendix~\ref{app:mergin_and_exact_GHD} for additional detail.
\par
\begin{figure}
    \centering
    \includegraphics[width=\linewidth]{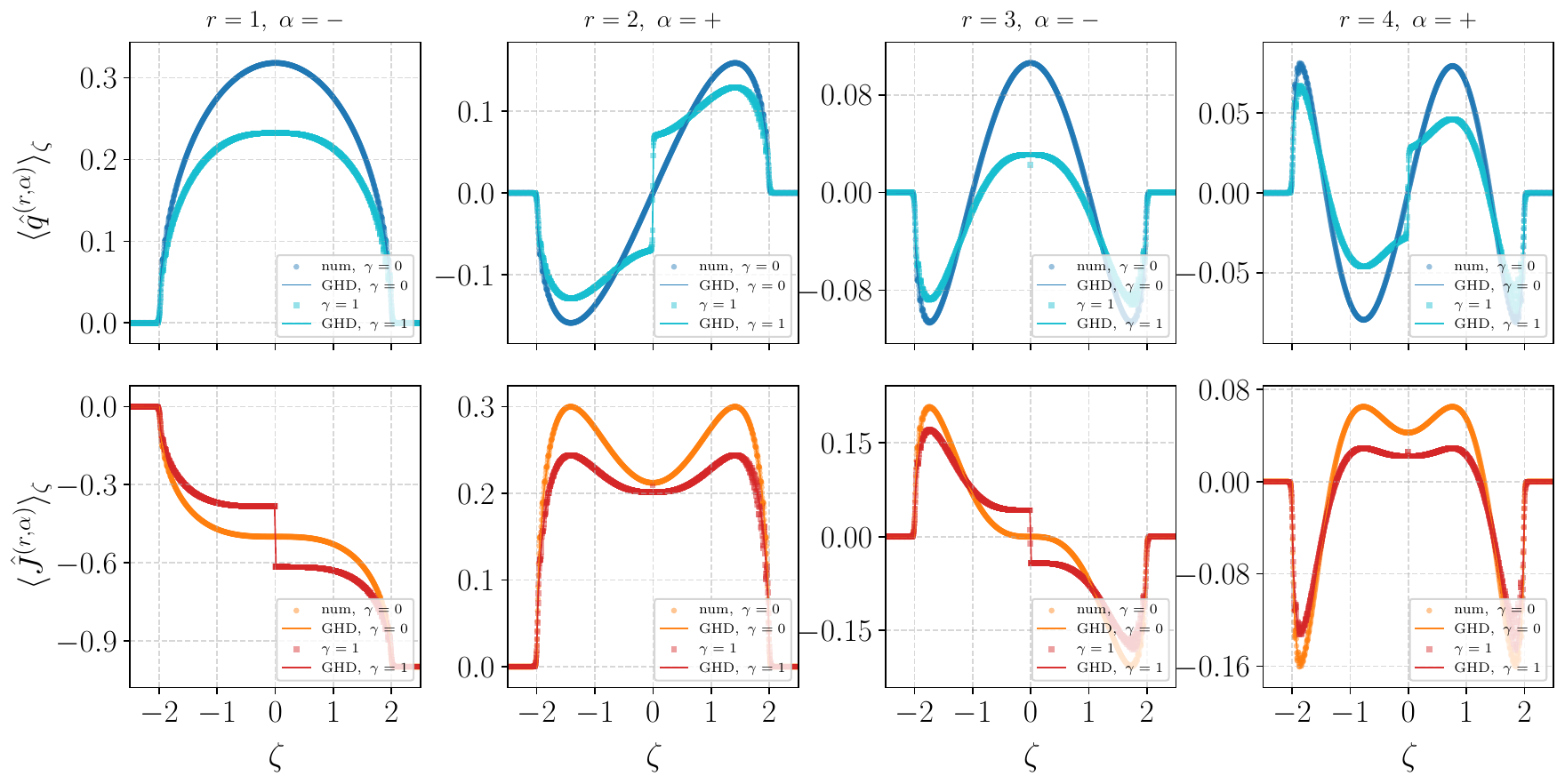}
    \caption{Local charge and current density profiles at late times for the initial domain-wall state. The first and second rows show charge and current profiles, respectively. We set $ J=1, \gamma={0, 1}, L=2N=4000$ sites,  $tJ=250$ and correction cutoff $r_{\rm cut}=5$. Local density profiles predictions using our GHD approach match exactly numerical results.}
    \label{fig:4_DW}
\end{figure}

We begin by taking an appropriate ansatz for the unknown root density 
\begin{align}
    \chi^L(k) &= n^R(k) + \phi^L(k)= \frac{1}{4\pi} +\phi^L(k), \label{eq:chiL} \\[5pt]
    \chi^R(k) &= n^L(k) + \phi^R(k)= \phi^R(k) \label{eq:chiR} .
\end{align}
Note that this choice is natural, since it reproduces the standard result for the GGE root density in the absence of measurements. In the above expressions, $n^{R}(k)=1/(4\pi)$ and $n^L(k)=0$ are the root densities of the vacuum-N\'eel bipartition defined in Eqs. \eqref{eq:initial_state_DW}, \eqref{eq:initial_vacuum-neel}. Next, we proceed by inserting the ansatz into the merging conditions ~\eqref{eq:cond1} and~\eqref{eq:cond2}.  As shown in Appendix~\ref{app:mergin_and_exact_GHD},
this yields the following two equalities
\begin{equation}
     \int^\pi_{-\pi} dk \sin(rk) \phi^-(k) = 0, \label{eq:parityM} 
\end{equation}
\begin{align}
 -4J^2 \int^\pi_{-\pi} dk \; \sin(rk ) \sin(k) \phi^-(k)   - 2 r\gamma  J\int^{\pi}_0 dk \sin(rk)\phi^R(k) = - \frac{\gamma J}{2\pi} (1-(-1)^r) +\delta_r \ ,  \label{eq:jump_cond}
\end{align}
where we defined $\phi^-(k)= \Theta(k) \phi^R(k)- \Theta(-k) \phi^L(k)$. Since Eq.~\eqref{eq:parityM} fixes the parity of $\phi^-(k)$, we can take the following Fourier mode expansion for $\phi^R(k)$ 
\begin{equation}
    \phi^R(k)= \sum_{n=1}^\infty b_n(\gamma) \sin(nk). \label{eq:phiR}
\end{equation}
Lastly, inserting Eq.~\eqref{eq:phiR} in Eq.~\eqref{eq:jump_cond}, standard calculations yield the following equation for the Fourier coefficients $b_n(\gamma)$
\begin{align}
\sum^\infty_{n=1} b_n(\gamma)\left[ \pi r \gamma J \delta_{n,r} - f^{(n)}_r  \right]= \frac{ \gamma J(1- (-1)^r)}{2 \pi } -\delta_r\ ,   \label{eq:b_nMatrix}
\end{align}
where 
\begin{equation}
    f^{(n)}_r= - 8J^2 \int^\pi_0 dk \sin(rk) \sin(k) \sin(nk)\,. 
\end{equation}
This is a linear system for the coefficients $b_n(\gamma)$, which can be easily solved numerically by truncating the infinite sum to a certain value of  $n=N_{\rm max}$.  Once the coefficients $b_n(\gamma)$ are known, one has a full solution for the function $\chi(k)$. In Fig.~\ref{fig:4_DW}, we test its validity, by comparing the corresponding profiles with those obtained by a solution to Eq.~\eqref{eq:ODE_C_ij_explicit}. The data are obtained setting $N_{\rm max}=800$ with cutoff $r_{\rm cut}=5$. We note that the plots show excellent agreement for the profiles of all local charges and currents that we have considered. 

\begin{figure}
    \centering
    \includegraphics[height=0.33\textheight,width=0.66\textwidth]{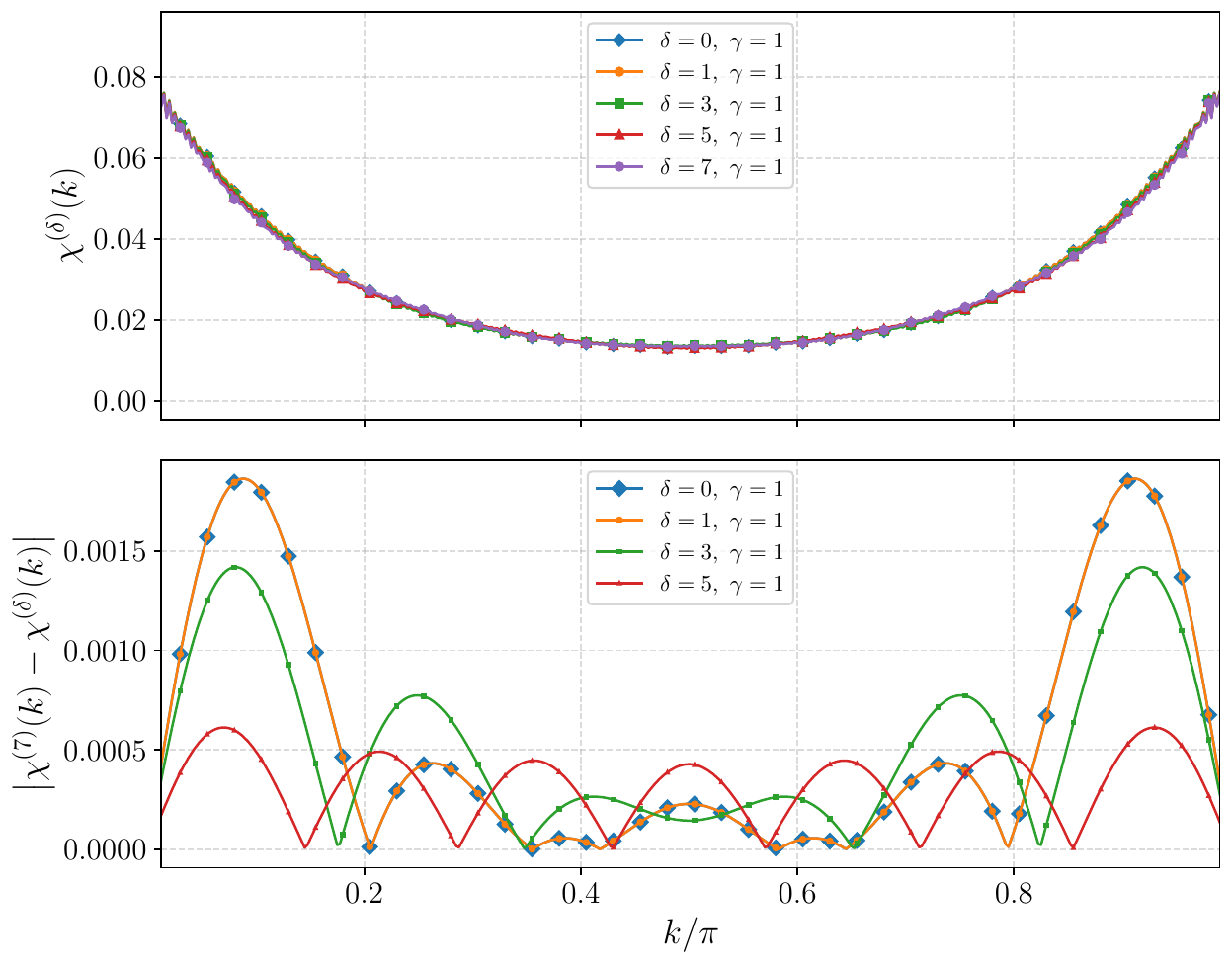}
    \caption{Upper panel: Root density profile for different cutoff values $r_{\rm cut}$. Lower panel: 
    plot of $|\chi^{(7)}(k)-\chi^{(\delta)}(k)|$ as a function of $k$, for increasing $\delta$.
    We set $J=1, \gamma={1}, N_{\rm max}=800$. The root density is slightly modified by inserting larger corrections. We note that, as $\delta$ increases, the maximum value of the difference $|\chi^{(7)}(k)-\chi^{(\delta)}(k)|$ decreases.}
    \label{fig:5_chiR}
\end{figure}

We note that discontinuities in the currents and charges appear only for the $\alpha=-$ and $\alpha=+$ case, respectively. In the case of  $q^{(1,-)}$, for instance, the absence of discontinuities reflects continuity of particle transport, since $\partial_t n_x = - (J^{(0,-)}_x - J^{(0,-)}_{x-1} )=  J(q^{(1,-)}_x - q^{(1,-)}_{x-1} )$. This observation is natural, since increasing the monitoring rate does not insert or deplete the system from particles.

\ \\
We now discuss the accuracy of the solution as a function of the cutoff $r_{\rm cut}$. In Fig.~\ref{fig:5_chiR} we report $\chi^{(r_{\rm cut})}(k)$ for increasing values of $r_{\rm cut}$. We see that the solution for $r_{\rm cut}=0$ already provides a very good approximation. This is also apparent from the study of the profiles for increasing $r_{\rm cut}$, cf. Appendix~\ref{sec:more_plots}.
\ \\

Besides the numerical solution, we can also study Eq.~\eqref{eq:b_nMatrix} in the limit of large monitoring. Dividing Eq.~\eqref{eq:b_nMatrix} by $\gamma$, and taking $\gamma\to\infty$, we obtain
\begin{equation}
    b_n(\gamma)= \begin{cases}
        \frac{1}{\pi^2 n}& n  \ {\rm odd}\\[3pt]
        0 & {\rm else}
    \end{cases} \quad ,
\end{equation}
where we assumed that the numerical correction $\delta_r $ scales as $\mathcal{O}(1)$ in $\gamma$.
Performing the infinite sum, we obtain
\begin{align}
    \phi^R(k)= \frac{1}{2\pi^2}\sum_{n=1}^\infty \frac{  (1- (-1)^n)}{ n }= \frac{1}{4\pi}, \label{eq:zeno_phi}
\end{align}
where we used the fact that the series on the right hand side of the equality sums to $\pi/2$. Hence, in the Zeno-limit, the root density $\chi(k)=\chi_\gamma(k)$ is given by
\begin{equation}
\lim_{\gamma \to \infty}\chi_\gamma^L(k) = 0, \quad \lim_{\gamma \to \infty}\chi_\gamma^R(k)=\frac{1}{4\pi},
\end{equation}
which coincides with the initial state root density, reflecting the freezing of the system due to the absence of transport. Finally, we comment that following a similar protocol, we can produce accurate predictions for the density profiles starting from the fully polarized state Eq.~\eqref{eq:fully_polarized}. Details of this calculation and numerical results are provided in Appendix~\ref{sec:more_general_states}

\ \\

% {\color{red} Add analytic derivation showing the Zeno-limit result for $\gamma\to\infty$}.

\section{GHD Solution from  homogeneous thermal states}
\label{sec:thermal_states}

In this section, we study the quench from an homogeneous thermal state
\begin{equation}\label{eq:thermal_state}
    \hat \rho_0=\frac{e^{-\beta \hat H}}{\mathcal{Z}}\,,
\end{equation}
where $\hat H$ is the Hamiltonian~\eqref{eq:Hamiltonian}, $\beta$ is the inverse temperature, while $\mathcal{Z}$ is a normalization constant. Note that, different from the initial state~\eqref{eq:initial_state}, $\hat \rho_0$ is an equilibrium state for the unitary dynamics, but the monitoring introduces a non-trivial evolution.

\subsection{The numerical solution}
\label{sec:sub_numerics_beta}
The initial conditions for Eq.~\eqref{eq:ODE_C_ij_explicit} corresponding to Eq.~\eqref{eq:thermal_state} can be easily computed based on the exact solution of the Hamiltonian~\eqref{eq:Hamiltonian}. In particular, we have 
% {\color{red} add formula}
\begin{equation}\label{eq:thermal}
    C_{x,y}(0)=\frac{1}{L}\sum_{m=-N}^{N-1} e^{i k_m (y-x)} \frac{1}{1+e^{-2J \beta \cos(k_m) }} \;,
\end{equation}
where we consider periodic boundary conditions for convenience and define the wave-number $k_m= (2\pi m)/L$, where we set $L=2N$. We have solved Eq.~\eqref{eq:ODE_C_ij_explicit} with initial conditions~\eqref{eq:thermal} for different values of $\beta$ and obtained the profiles for local charges and currents at late times.

\begin{figure}[H]
    \centering
    \includegraphics[width=\linewidth]{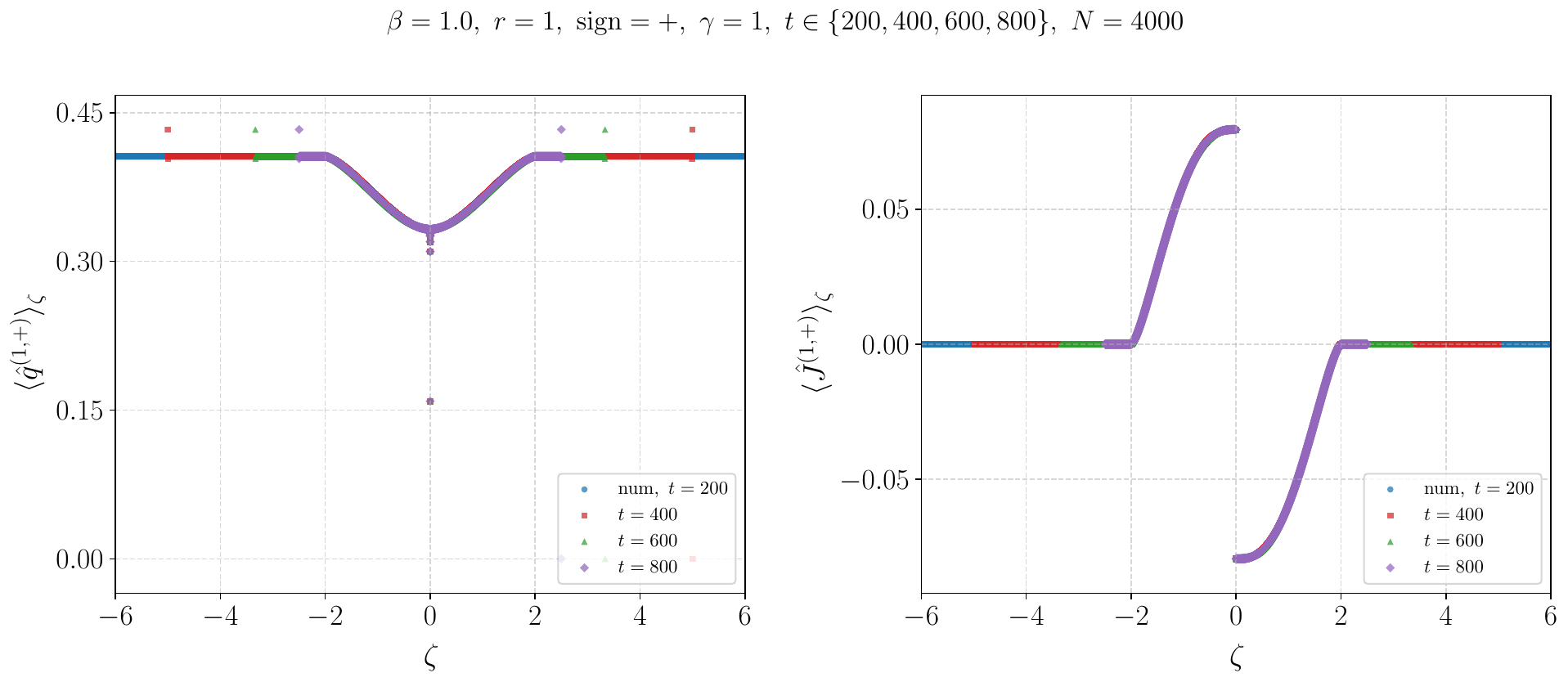}
    \caption{Charge and current profiles at the hydrodynamic scale for an initial homogeneous thermal state. We take the rescaling $\zeta=x/t$ for plotting profiles. a) Late time profiles of the local charge $\langle \hat{q}_x^{(1,+)}(t) \rangle$. The charge profile is continuous for all ray values. b) Late time profiles of the local current $\langle \hat{J}_x^{(1,+)}(t) \rangle$. Far from the propagation front ray $\zeta^*$, the current remains constant and equal to its initial value, displaying a discontinuity at $\zeta=0$. We set $\beta= 1, J=1, \gamma=1, L=2N=4000$ sites, and $tJ= \{200 ,400, 600, 800 \}$.
    }
    \label{fig:6}
\end{figure}

An example of our numerical data is reported in Fig.~\ref{fig:6}. The plots are reported in terms of the rescaled variable $\zeta=x/t$, and show a clear collapse of the profiles as time increases. We see that, perhaps counterintuitively, the monitoring causes a non-trivial variation of the profiles at hydrodynamic scales. In particular, we observe that the energy current develops a discontinuity near the origin. Physically, although the charge is conserved in the bulk of the region $A$, which implies absence of particle transport $\partial_t\langle \hat q^{(0,+)}\rangle =0$, the monitoring induces an effective defect at $\zeta=0$. In turn, the defect modifies the quasi-particle passing through the origin, changing the profiles at the hydrodynamic scales.

As in the domain-wall state case, we observe that the profiles of some charges are identically vanishing. However,  the vanishing profiles are the ones that  were non-zero in the domain-wall case.  This can be explained by the presence of a different unitary symmetry defined by the operator $\hat{V}$, such that $\hat{V} \hat{c}_j\hat{V}^{-1}= (-1)^j \hat{c}^\dagger_j.$ This unitary transformation (a \lq\lq staggered\rq\rq particle-hole exchange) leaves the Hamiltonian invariant and adds a constant term to the partial charge $\hat{Q}_A$ since
\begin{align}
     &\hat{V} \hat{c}^\dagger_{j+1}\hat{c}_j\hat{V}^{-1}= -\hat{c}_{j+1} \hat{c}_{j}^\dagger = \hat{c}_{j}^\dagger\hat{c}_{j+1},\\
     &\hat{V} \hat{c}^\dagger_{j}\hat{c}_j\hat{V}^{-1}= 1- \hat{c}^\dagger_{j}\hat{c}_j.
\end{align}
Plugging this into the Lindbladian, we find that
\begin{align}
    \hat{V }\mathcal{L}(\hat{\rho})\hat{V}^{-1}= \mathcal{L}(\hat{V }\hat{\rho}\hat{V}^{-1})\ .\label{eq:ph_symm}
\end{align}
Since $\hat{V}$ commutes with the initial thermal state, from the explicit expressions Eqs. \eqref{eq:qp}, \eqref{eq:qm}, we obtain
\begin{align}
   &\langle \hat{V} \hat{q}^{(2r+1,-)}_x\hat{V}^{-1} \rangle = -\langle \hat{q}^{(2r+1,-)}_x \rangle ,\\
    &\langle \hat{V} \hat{q}^{(2r,+)}_x\hat{V}^{-1} \rangle = -\langle \hat{q}^{(2r,+)}_x \rangle,
\end{align}
therefore
\begin{equation}
    \langle \hat{q}^{(2r+1,-)}_x \rangle = \langle \hat{q}^{(2r,+)}_x \rangle=0\ .
\end{equation}

We also note that this quench displays an additional symmetry. That is, spatial inversion or parity with respect to the bond between two neighboring sites. Namely, letting $\hat{\mathcal{P}}\hat{c}_j\hat{\mathcal{P}}^{-1}= \hat{c}_{l-j}$, the Hamiltonian is invariant for every choice of $l$ due to periodic boundary conditions, $\hat{\mathcal{P}}\hat{H}\hat{\mathcal{P}}^{-1}= \hat{H}$.
By picking $l=1$ (i.e. the bond between sites 0 and +1), then\footnote{Strictly speaking, Eq. \eqref{eq:parity_symmetry_Q_A} holds exactly for $A=\{1,\dots,N-1\}$ and $\bar{A}=\{-N+2,\dots,0\}$, whereas according to the convention outlined in Sec. \ref{sec:the_protocol}, $\bar{A}=\{-N,\dots,0\}$. However, the reflection around the bond between $\bar{A}$ and $A$ becomes an exact symmetry in the thermodynamic limit $N\to \infty$.}
\begin{equation}
\label{eq:parity_symmetry_Q_A}
\hat{\mathcal{P}}\hat{Q}_A\hat{\mathcal{P}}^{-1}=\hat{Q}_{\bar{A}} \equiv \hat{Q} - \hat{Q}_{A}.
\end{equation}
Hence, we find that the Lindbladian transforms as
\begin{equation}
    \hat{\mathcal{P} }\mathcal{L}(\hat{\rho})\hat{\mathcal{P}}^{-1}= 
    \mathcal{L}(\hat{\mathcal{P} }\hat{\rho}\hat{\mathcal{P}}^{-1})\ .\label{eq:parity_symm}
\end{equation}
Since the initial state is $U(1)$-invariant, at any time the evolved state is invariant under the action of $\hat{\mathcal{P}}$. Thus, we can combine the action of $\hat{V}$ and $\hat{\mathcal{P}}$ to show that
\begin{align}
    \langle \hat{q}^{(2r,-)}_{x} \rangle_t=-\langle \hat{q}^{(2r,-)}_{-x} \rangle_t\;, \quad \langle \hat{q}^{(2r+1,+)}_{x} \rangle_t= \langle \hat{q}^{(2r+1,+)}_{-x} \rangle_t. \label{eq:invsymm_q}
\end{align}

In the next subsection, we provide an analytic solution to the monitored dynamics following the general strategy outlined in 
Sec.\ref{sec:ghd_monitored_dyn_theory}. In this case, we implement the merging conditions using the boundary term $\mathcal{Q}(r,\pm)$ as defined in Eq. \eqref{eq:Q_bdy}, without making any ansatz on its structure.
\subsection{Merging conditions and the GHD solution}
\begin{figure}[H]
    \centering
    \includegraphics[width=\linewidth]{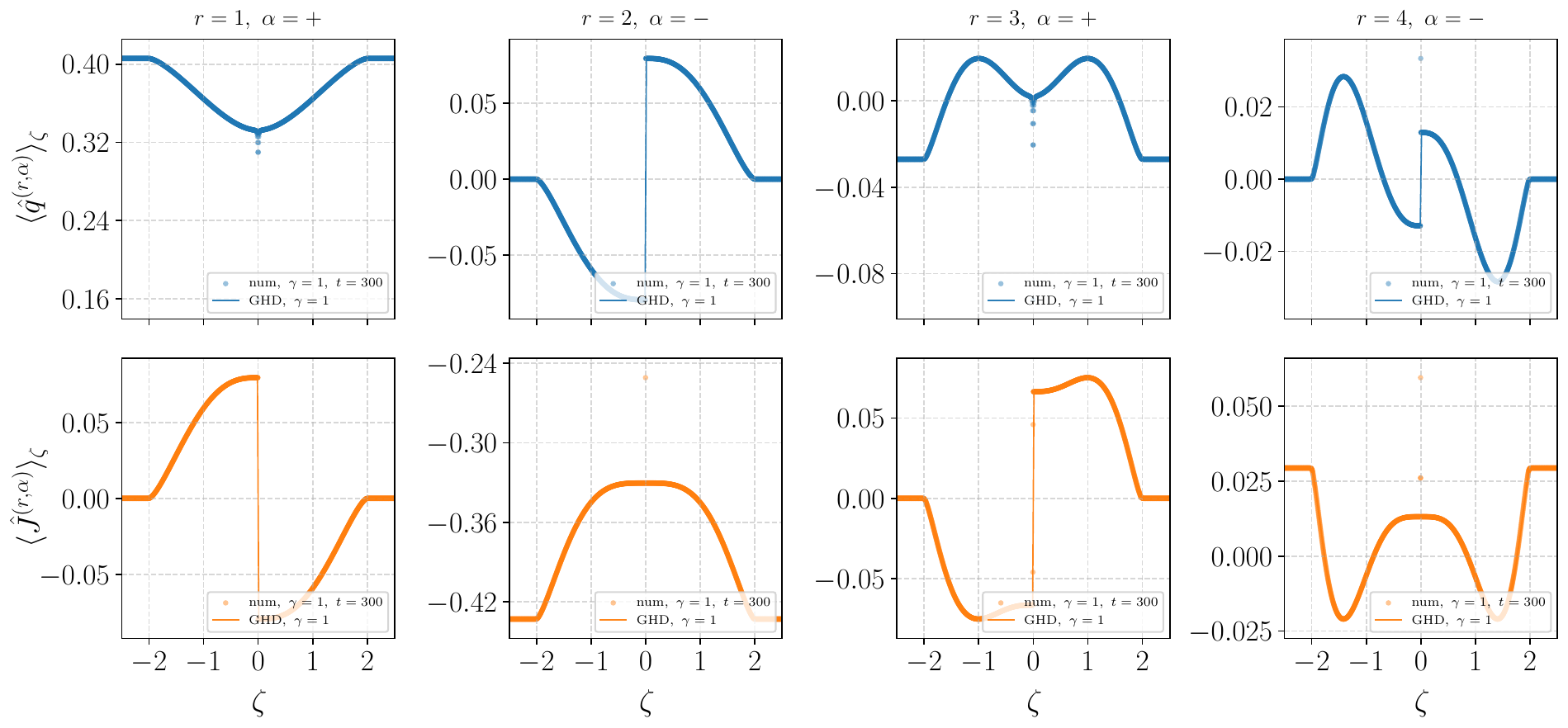}
    \caption{Charge and current profiles at the hydrodynamic scale for an initial homogeneous thermal state. The first and second rows show the charge and current profiles. We set $\beta= 1, J=1, \gamma=1, L=2N=4000$ sites, $tJ= 300$ and cutoff $r_{\rm cut}=7$. Local density profiles predictions using our GHD approach are in perfect agreement with the numerical results.}
    \label{fig:7_GHDbeta}
\end{figure}

Starting from the symmetry constraint stemming from the Lindbladian inversion symmetry on the local profiles, Eq.~\eqref{eq:invsymm_q} tells us that in the hydrodynamic regime
\begin{align}
      \lim_{x \to \infty} \langle \hat q_x^{(2r+1,+)}(\infty)\rangle &=\lim_{x \to\infty} \langle \hat q_{-x}^{(2r+1,+)}(\infty)\rangle \nonumber \\
      \Rightarrow
      \langle \hat q^{(2r+1,+)}\rangle_{\zeta=0^+} &=\langle \hat q^{(2r+1,+)}\rangle_{\zeta=0^-}. \label{eq:beta_cond1}
\end{align}
In this case, we were not able to provide a GHD ansatz for the boundary terms $\mathcal{Q}{(r,+)}$. Instead we proceeded by supplementing the merging condition Eq.~\ref{eq:mergin_equations} with data from long time simulations for $N=4000$ and $t=300$. We use the following ansatz for the unknown root density 
\begin{align}
    \chi^L(k) &= n^\beta(k) + \phi^L(k)\, \quad 
    \chi^R(k) = n^\beta(k) + \phi^R(k)\,,
\end{align}
where
\begin{equation}
n^\beta(k) =\frac{1}{1+e^{-2J\beta \cos(k)}}
\end{equation}
is the root density of the initial thermal state. Inserting the ansatz into the merging conditions~\eqref{eq:mergin_equations} and~\eqref{eq:beta_cond1}, we obtain two equalities
\begin{equation}
    \int^\pi_{-\pi} dk \cos(rk)\phi^-(k) = 0 ,\label{eq:parityM_beta} 
\end{equation}
\begin{equation}
    4J^2 \int^\pi_{-\pi} dk \; \cos(rk ) \sin(k) \phi^-(k)   = - \gamma \mathcal{Q}^{(r,+)}(\infty) .\label{eq:jump_cond_beta}
\end{equation}
where $\phi^-(k)$ is defined as in Sec. \ref{sec:sub_ghd_analytic}. Since Eq.~\eqref{eq:parityM_beta} fixes the parity of $\phi^-(k)$, we can take the following Fourier mode expansion for $\phi^R(k)$ 
\begin{equation}
    \phi^R(k)= \sum_{n=1}^\infty a_n(\gamma) \cos(nk). \label{eq:phiR_2}
\end{equation}
Lastly, inserting Eq.~\eqref{eq:phiR_2} in Eq.~\eqref{eq:jump_cond_beta}, standard calculations yield the following equation for the Fourier coefficients $a_n(\gamma)$
\begin{align}
\sum^\infty_{n=1} a_n(\gamma) g^{(n)}_r  =  \mathcal{Q}^{(r,+)}(\infty) ,   \label{eq:a_nMatrix}
\end{align}
where 
\begin{equation}
    g^{(n)}_r=  8J^2 \int^\pi_0 dk \cos(rk) \sin(k) \cos(nk) \label{eq:f_n_r}. 
\end{equation}
As for the domain-wall case, this is a linear system for the coefficients $a_n(\gamma)$, which can be easily solved numerically by truncating the infinite sum to a certain value of  $n=N_{\rm max}$. In Fig.~\ref{fig:7_GHDbeta}, we test the validity of our GHD prediction, by comparing the corresponding profiles with those obtained by a solution to Eq.~\eqref{eq:ODE_C_ij_explicit}. Data are obtained by setting $N_{\rm max}=800$ with cutoff $r_{\rm cut}=7$. We note that plots show excellent agreement for the profiles of all local charges and currents that we have considered. 

\section{Conclusions}
\label{sec:conclusions}

We have developed a GHD approach to describe the late time dynamics of free fermionic systems  subject to continuous monitoring of a charge over an extensive region. We have shown the emergence of ballistic propagation of quasi-particles, in contrast to the well known diffusive behavior of free fermions subject to local-density monitoring~\cite{medvedyeva2016Exact,IshiyamaFujimotoSasamoto2025ExactDensityProfile,znidaric2010exact,znidaric2014exact}. Within the bipartition protocol, we have provided an hybrid numerical-analytic solution of the quench dynamics from some special initial states. Our solution predicts discontinuities in the profiles of local charges and currents, that become more pronounced as the monitoring rate increases and that lead to absence of transport in the Zeno limit of infinite monitoring rate.

Our work opens several directions for future research. First, it would be  important to develop a systematic approach to derive the merging conditions for arbitrary initial states. Going further, a very natural direction would be to extend our approach to the case of interacting systems. In this respect, we emphasize that, being based on the GHD, our approach is naturally tailored to be extended to interacting systems within the (thermodynamic) Bethe ansatz formalism~\cite{takahashi1999thermodynamics}. The main difficulty here is technical rather than conceptual since we need to feed our ansatz with microscopic information. In the free case, we obtained this information from the two-point correlation functions, whereas in the integrable (interacting) case, conserved charges have more complex expressions, and obtaining their expectation values at hydrodynamic scales involve solving thermodynamic Bethe ansatz equations.

Finally, since the total particle number can be efficiently measured in quantum-circuit setups~\cite{buhrman2024state,piroli2024approximating,rethinasamy2024logarithmic,zi2025shallow}, a natural question pertains to the extension of our results to discrete settings, which are relevant for the implementation of integrable circuits in current digital quantum platforms~\cite{morvan2022formation,maruyoshi2023conserved,keenan2023evidence}. We also note that current cold-atom measurement platforms could be potentially relevant for the realization of our model, since dispersive and cavity-assisted detection schemes enable weak measurements over extended regions~\cite{mekhov2009quantum}. The main bottleneck for realizing our proposed measuring protocol lies in the difficulties of keeping measurement-induced heating and decoherence sufficiently weak to reach hydrodynamic scales~\cite{altuntas2023weak}.
A natural question pertains to the extension of our results to discrete settings, which are relevant for the implementation of integrable circuits in current digital quantum platforms~\cite{morvan2022formation,maruyoshi2023conserved,keenan2023evidence}. Indeed, integrable quantum circuits have been shown to display interesting non-equilibrium features that go beyond those of traditional Hamiltonian models~\cite{vanicat2018integrable,ljubotina2019ballistic,aleiner2021bethe,giudice2022temporal,claeys2022correlations,miao2023integrable,vernier2023integrable,vernier2024strong,hubner2025generalized,paletta2025integrability,paletta2025integrability_2} and that can also be captured by GHD~\cite{hubner2025generalized}. We leave these directions for future work.

\label{sec:conclusion}

\section*{Acknowledgements}
 We thank Bruno Bertini, Luca Capizzi, Federico Carollo, Maurizio Fagotti, Stefano Scopa, and Riccardo Travaglino for useful discussions.

\paragraph{Funding information}
This work was funded by the European
Union (ERC, QUANTHEM, 101114881). Views and opinions expressed are however those of the
author(s) only and do not necessarily reflect those of the European Union or the European Research Council Executive Agency. Neither the European Union nor the granting authority can be held responsible for them.

\begin{appendix}

\section{Master Equation Derivation}\label{App.Master}

In this appendix, we derive the equation of motion for the monitored dynamics. Because we are interested in the averaged dynamics, we will only derive the Lindbladian description. We emphasize, however, that one could also write down a stochastic equation depending on the (random) measurement outcomes and describing individual quantum trajectories~\cite{diosi1998non,caves1987quantum}. 

Recall that we are interested in the charge operator
\begin{equation}
    \hat{Q}_A= \sum_{j\in A} \hat{n}_j = \sum_{m=0}^{|A|} m\; \hat{\Pi}^{(A)}_m ,
\end{equation}
where $\hat{\Pi}_m^{(A)} =(\hat\Pi_m^{(A)})^\dagger= (\hat\Pi_m^{(A)})^2$ is a projector onto the eigenspace of $\hat{Q}_A$ with eigenvalue $m$ and $\sum_m \hat{\Pi}^{(A)}_m = \hat{1}$. Let us denote by $\hat{\rho}(t)$ the state of the system at time $t$. Following the protocol explained in Sec.~\ref{sec:the_protocol}, we first let our system evolve unitarily up to time $t+\Delta t$
\begin{equation}
    \hat\rho(t)\mapsto \hat\rho(t+\Delta t)=e^{-i\hat{H}\Delta t}   \hat \rho(t)e^{i\hat{H}\Delta t}\,.
\end{equation}
Then, with probability $p=\gamma \Delta t$ we measure the charge $\hat{Q}_A$.  Accordingly, the averaged density matrix is transformed as
\begin{equation}
\hat\rho(t+\Delta t)\mapsto \hat\rho'(t+\Delta t)=(1-\gamma \Delta t) \hat\rho (t+\Delta t)+\gamma \Delta t \sum_{m}\hat\Pi_m^{(A)} \hat\rho(t+\Delta t) \hat\Pi_m^{(A)} \,.
\label{DeltaMasterEQ}
\end{equation}
 Expanding to the first order in $\Delta t$ and taking the limit $\Delta t\to 0$ yields the following GKSL equation~\cite{breuer2002theory}
 
\begin{equation}
     \partial_t \hat{\rho}(t)= -i[\hat{H},\hat{\rho}(t)] +\gamma \sum_m \left(\hat{\Pi}^{(A)}_m \hat{\rho}(t)\hat{\Pi}^{(A)}_m - \frac{1}{2}\left\{ \hat{\Pi}^{(A)}_m\hat{\Pi}^{(A)}_m,\hat{\rho}(t) \right\}\right)\,,
     \label{MasterEQGKSL}
\end{equation}
 which ensures the dynamics studied is CPTP. Using $\hat{\Pi}^{(A)}_m\hat{\Pi}^{(A)}_m=\hat{\Pi}^{(A)}_m,\; \sum_m \hat{\Pi}^{(A)}_m = \hat{1}$, we obtain

 \begin{equation}
     \partial_t \hat{\rho}(t)= -i[\hat{H},\hat{\rho}(t)] +\gamma \left(\sum_m \hat{\Pi}^{(A)}_m \hat{\rho}(t)\hat{\Pi}^{(A)}_m - \hat{\rho}(t)\right)\,.
     \label{MasterEQ}
\end{equation}
 
Finally, expanding $\hat \rho(t)$ into eigenstates of the charge operator $\hat{Q}$, it is immediate to show 
\begin{equation}
    \sum_m \hat{\Pi}^{(A)}_m \hat{\rho}(t)\hat{\Pi}^{(A)}_m =\frac{1}{2\pi} \int_0^{2\pi} d\alpha e^{i\alpha \hat{Q}_A} \hat{\rho}  e^{-i\alpha \hat{Q}_A} \,,
\end{equation}
which concludes the derivation of Eq.~\eqref{eq:Lindblad}.
% \textcolor{blue}{We conclude this appendix by showing that the equation of motion for the reduced density matrix is indeed a CPTP map. The trace preserving property follows from:
% \begin{equation}
%     \partial_t {\rm Tr}  [\hat{\rho}(t)] = -i{\rm Tr} [\hat H, \hat \rho(t)]  +\gamma {\rm Tr}[\sum_m \hat\Pi_m^{(A)} \hat \rho(t)- \hat\rho(t)]  = 0,
% \end{equation}
% where we used the completeness of the charge projectors $\sum_m \Pi_m^{(A)} =I_A$, and the fact that the trace of a commutator vanishes.The complete positiveness of the map can be shown by bringing 
% \begin{equation}
% \hat\rho(t+\Delta t)\mapsto \hat\rho'(t+\Delta t)=(1-\gamma \Delta t) \hat\rho (t+\Delta t)+\gamma \Delta t \sum_{m}\hat\Pi_m^{(A)} \hat\rho(t+\Delta t) \hat\Pi_m^{(A)} \,,
% \label{DeltaMasterEQ}
% \end{equation}
% into Krauss form. Defining $\hat M_0 = I -i \hat H dt - dt\frac{\gamma}{2}\sum_m \hat \Pi_m^{(A)}$ and $\hat M_{m+1} = \sqrt{\gamma dt}\; \hat \Pi_m^{(A)}$, for $m=0,\dots, |A|$, a simple calculation yields 
% \begin{equation}
%     \rho(t+\Delta t) = \sum_m \hat M_m \hat \rho(t) \hat M^\dagger_m\,,
% \end{equation}
% up to terms of order $\mathcal{O}(dt^2)$, and $\sum_m \hat M_m M^\dagger_m =I +\mathcal{O}(dt^2)$.  The Krauss form of a map is complete positive, therefore, we conclude that our master equation is indeed a CPTP map.}
% % % % % % % % % % % %  % % % % % % % % % % % % 

\section{Technical details}
\label{Appx:Identities}

In this appendix we provide details on the technical calculations reported in Sec.~\ref{sec:micro_general}

\subsection{Details on the numerical calculations}
\label{sec:numeric_micro_extraction}

In this section, we provide more details on the extraction of the microscopical data necessary to produce accurate profiles at large times. Using sparse matrix methods and solving  Eq.~\eqref{eq:ODE_C_ij_explicit} using a 4 point Runge-Kutta algorithm, we were able to produce numerics of systems sizes up to $N=10000$. Focusing on $N=4000$, we extracted the values of the correction vector $\delta_r$ for the N\'eel and fully polarized states as follows.

First, we computed numerically an estimate for $\mathcal{Q}(r,-)$, for $r\leq r_{\rm cut}$. We did this by running the numerical simulations up to times at which $\mathcal{Q}(r,-)$ appeared converged up to a given numerical precision. Next, we estimated numerically $\langle \hat q^{(r,-)}\rangle_{\zeta=0}$ via the identification~\eqref{eq:zeta0lim}, by extracting  $\langle \hat q^{(r,-)}_x(\infty)\rangle$ for $x$ sufficiently far away from the origin. As it is apparent from Fig.~\ref{fig:NESS_profiles}, the values of $\langle \hat q^{(r,-)}_x(\infty)\rangle$ are essentially constant already for small values of $x$. Then, we computed 
\begin{equation}
    \delta_r=\mathcal{Q}(r,-)-r\langle \hat q^{(r,-)}\rangle_{\zeta=0}\,.
\end{equation}

The above procedure comes with two different sources of approximation errors. The first one is a finite-time effect, due to the fact that we estimate infinite-time quantities at finite values of $t$. The second one comes from estimating $\lim_{x\to\infty}\langle \hat q^{(r,-)}_{x}(\infty)\rangle$ in Eq.~\eqref{eq:zeta0lim} from finite values of $x$. We estimated that both approximation errors are very small for the time and distances chosen, introducing errors in the profiles that are invisible at the scales of our plots. 

Finally, we followed a similar strategy for the homogeneous thermal case, where we extract the values of $\mathcal{Q}{(r,+)}$ directly from the covariance matrix data up to the cutoff value $r_{\rm cut}$.

\subsection{The merging conditions and the exact GHD solution}
\label{app:mergin_and_exact_GHD}

In this section, we provide more details for the calculations required to obtain $\chi(k)$ in Sec.~\ref{sec:sub_ghd_analytic}. To derive Eqs.~\eqref{eq:parityM}, \eqref{eq:jump_cond}, note that, from Eq. \eqref{eq:sol} it follows that the average of a local observable $g$ over the GGE root density $n_\zeta(k)$ in the limit $\zeta \to 0^\pm$ is given by
\begin{align}
\langle g\rangle_{0^+} &= \int_0^\pi dk\, \chi^R(k) g(k) + \int_{-\pi}^0dk\,n^R(k) g(k)\,,\nonumber\\
\langle g\rangle_{ 0^-} &= \int_0^\pi dk\, n^L (k) g(k) + \int_{-\pi}^0dk\,\chi^L(k) g(k)\,. \label{eq:GGE_zeta0}
\end{align}
Inserting the ansatz for the root densities $\chi^{R/L}(k)$  into the merging conditions Eqs. \eqref{eq:cond1}, \eqref{eq:cond2} and using Eq.~\eqref{eq:GGE_zeta0}, with the function $g(k)$ read from Eqs. \eqref{eq:ghd_q} and \eqref{eq:ghd_j}, one immediately obtains Eqs.~\eqref{eq:parityM}--\eqref{eq:jump_cond}. Lastly, a simple calculation yields for $r\neq n$
\begin{align}
    f^{(n)}_r 
   &= 16 J^2 \frac{rn(1+(-1)^{r+n})}{n^4 -2n^2(r^2+1) + (r^2-1)^2
    }\,.
\end{align}
Similarly, for $r=n$
\begin{align}
    f^{(n)}_n & = 32J^2\frac{n^2}{1-4n^2}.
\end{align}
Finally, for the homogeneous thermal case, we find 
\begin{equation}
    g_r^{(n)}= 
        -8 J^2 \frac{(n^2+r^2-1)(1+(-1)^{r+n})}{n^4 -2n^2(r^2+1) + (r^2-1)^2}. 
\end{equation}

\subsection{Additional numerical results}
\label{sec:more_plots}

\begin{figure}
    \centering
    \includegraphics[width=\linewidth]{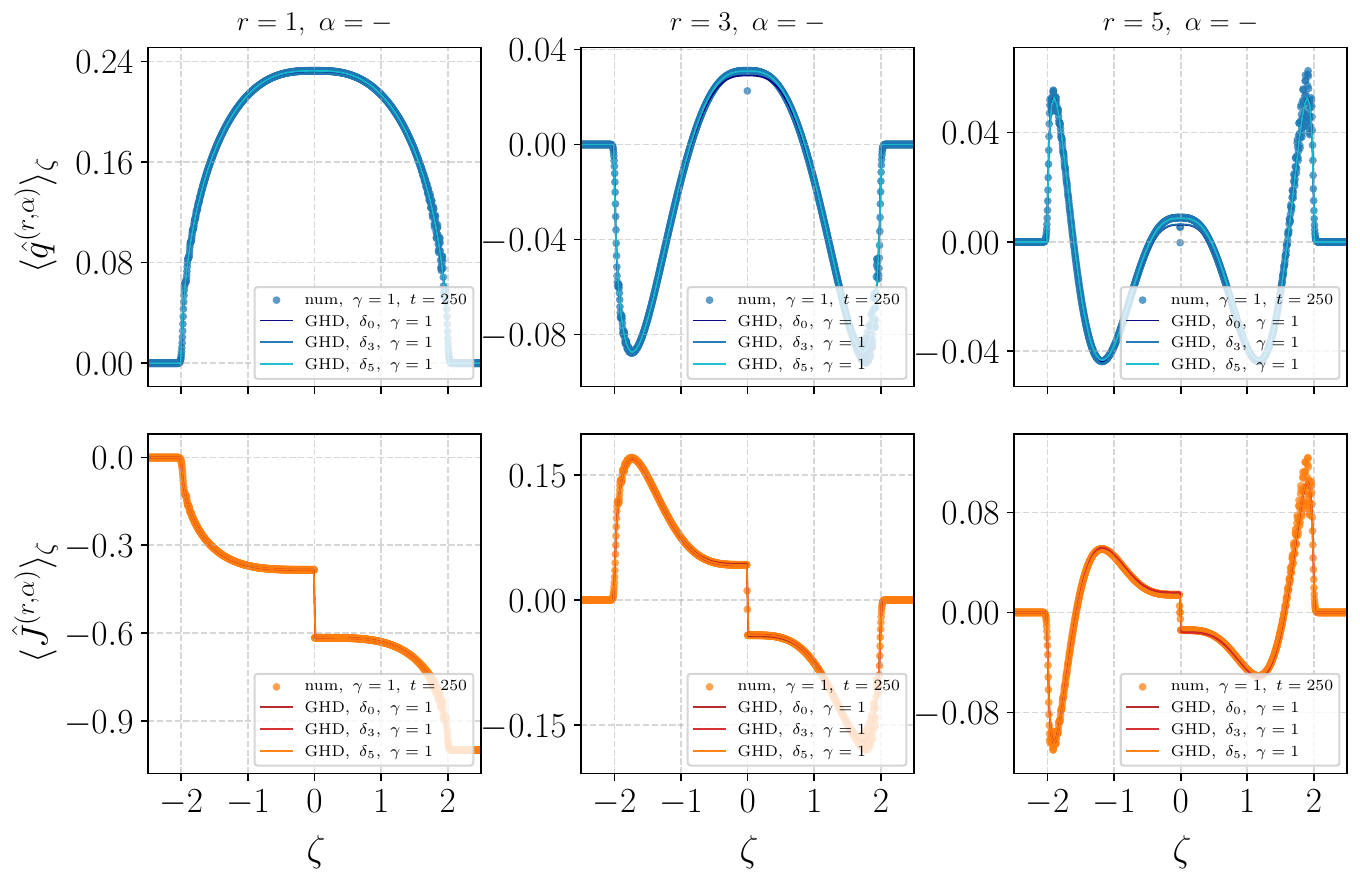}
    \caption{Local charge and current density profiles at late times for the initial domain-wall state. The first and second rows show the charge and current profiles, respectively. We set $ J=1, \gamma={0, 1}, L=2N=4000$ sites,  $tJ=250$ and cutoffs $r_{\rm cut}=\{0, 3, 5 \}$. The accuracy of our GHD predictions against numerics increases rapidly as we increase the cutoff value $r_{\rm cut}$. Note that the only noticeable difference between profiles lies within a small neighborhood around $\zeta=0$. Note also that finite-time effects are visible in the profiles for $r=5$, at large distances from the origin.}
    \label{fig:8_Rneel}
\end{figure}

In this section, we provide additional numerical evidence supporting the validity of our GHD solution for the domain-wall and homogeneous thermal state presented in the main text. Fig.~\ref{fig:8_Rneel} shows the profiles of local charges and currents for several values of $r$ for the domain-wall state case. We note that small corrections coming from microscopical details become apparent only for large values of $r$ and within a small neighborhood of the $\zeta=0$ ray.

\begin{figure}
    \centering
    \includegraphics[width=\linewidth]{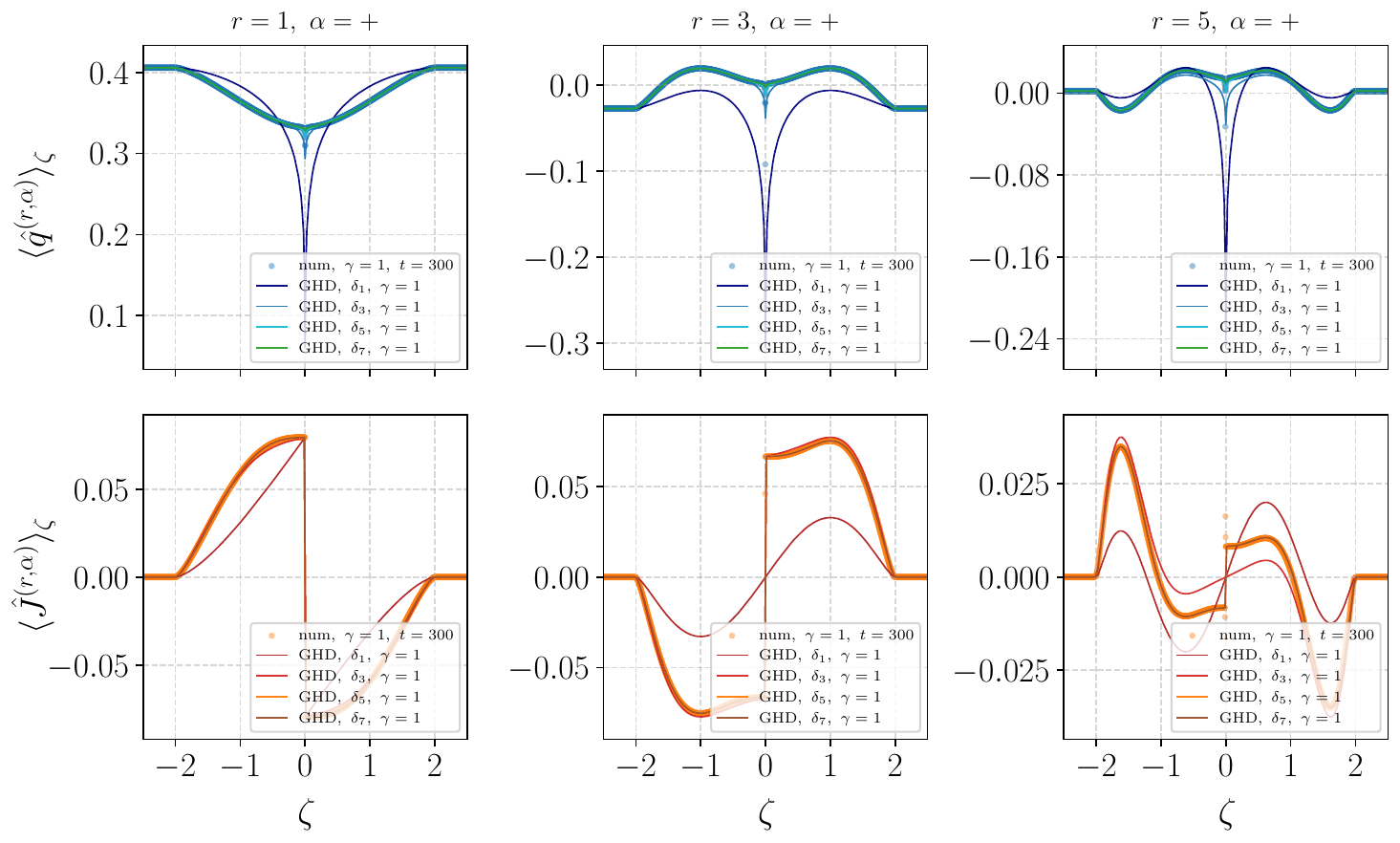}
    \caption{Local charge and current density profiles at late times for the initial homogeneous thermal state. The first and second rows show the charge and current profiles, respectively. We set $ J=1, \gamma=1, \beta=1, L=2N=4000$ sites,  $tJ=250$ and cutoffs $r_{\rm cut}=\{1, 3, 5, 7 \}$. Accuracy of our GHD profiles predictions against numerics increases rapidly as we increase the cutoff value $r_{\rm cut}$.}
    \label{fig:9_Rbeta}
\end{figure}

Next, Fig.~\ref{fig:9_Rbeta} shows the profiles of local charges and currents for several values of $r$ for the initial homogeneous thermal state case. We note that the accuracy of our predictions improves rapidly by introducing more $\mathcal{Q}{(r,+)}$ terms.

\section{GHD solution for the fully polarized state}
\label{sec:more_general_states}
\begin{figure}[H]
    \centering
    \includegraphics[width=\linewidth]{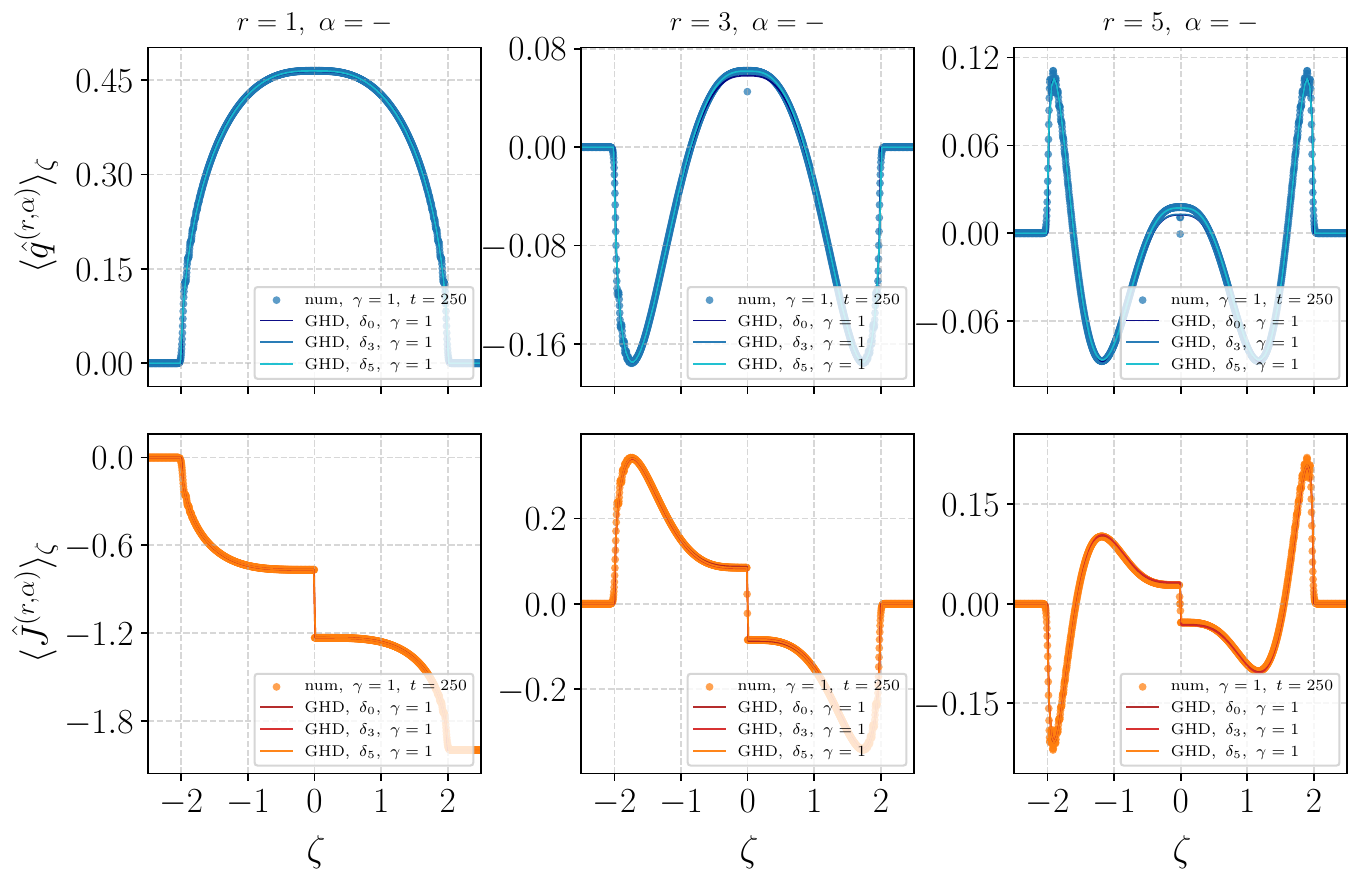}
    \caption{Local charge and current density profiles at late times for the initial fully-polarized state. The first and second rows show charge and current profiles, respectively. We set $ J=1, \gamma={ 1}, L=2N=4000$ sites, and $tJ=250$ and cutoffs $r_{\rm cut}=\{0, 3, 5 \}$. Accuracy of our GHD profiles predictions against numerics increases rapidly as we increase the cutoff value $r_{\rm cut}$. Note that the only noticeable difference between profiles lies within a small neighborhood around $\zeta=0$.}
    \label{fig:10_polarized}
\end{figure}

In this appendix we present the GHD solution for a system initially prepared in a fully polarized state Eq~\eqref{eq:fully_polarized}.  We begin by proving the boundary conditions that allow us to fix the unknown density $\chi(k)$. Using similar arguments as those described in the previous appendix, we obtain the following boundary conditions
\begin{align}
    \langle \hat{q}^{(r,-)}\rangle_{\zeta={0-}} = \langle \hat{q}^{(r,-)} \rangle_{\zeta={0+}} &= :\langle \hat{q}^{(r,-)} \rangle_{0}, \label{eq:cond1_polarized} \\
    \langle \hat{J}^{(r,-)} \rangle_{\zeta={0^+}} -\langle \hat{J}^{(r,-)} \rangle_{\zeta={0^-}}&=-\gamma r \langle \hat{q}^{(r,-)} \rangle_{0} +\delta_r , \label{eq:cond2_polarized}
\end{align}
which are the exact same boundary conditions as for the case of the N\'eel state. Since the only difference between the N\'eel and the fully-polarized state is the initial root density in the right half, the same procedure outlined in the main text and in the appendix~\ref{Appx:Identities} applies, taking into account that $n^R(k)= 1/2\pi$.
Hence, the matrix equation for the Fourier coefficients $b_n(\gamma)$ is given by  
\begin{align}
\sum^\infty_{n=1} b_n(\gamma)\left[ \pi r \gamma J \delta_{n,r} - f^{(n)}_r  \right]= \frac{ \gamma J(1- (-1)^r)}{ \pi } +\delta_r \ .   \label{eq:b_nMatrix_polarized}
\end{align}
Fig.~\ref{fig:10_polarized} shows excellent agreement between exact numerics and our GHD predictions for the late time profiles of several charge and current densities for several measuring rates. We truncate the infinite sum of Eq.~\eqref{eq:b_nMatrix_polarized} at $N_{\rm max}=800$ when solving for $b_n(\gamma)$.
\bibliography{bibliography}

\end{appendix}

\nolinenumbers

\end{document}